%% file: main.tex
\pdfoutput=1
\documentclass[10pt, conference, letterpaper]{IEEEtran}

\usepackage[utf8]{inputenc}

\input{Sections/control-wasiur}

\IEEEoverridecommandlockouts

\title{A Generalized Performance Evaluation Framework for  Parallel Systems with Output Synchronization\ifthenelse{\boolean{longVersion}}{\\(Extended Version)}{}
}

\author{\IEEEauthorblockN{Wasiur R. KhudaBukhsh\IEEEauthorrefmark{1}, Sounak Kar\IEEEauthorrefmark{2}, Amr Rizk\IEEEauthorrefmark{2},   and Heinz Koeppl\IEEEauthorrefmark{1}}
	\IEEEauthorblockA{\IEEEauthorrefmark{1}Bioinspired Communication Systems Lab (BCS),}
	\IEEEauthorblockA{\IEEEauthorrefmark{2}Multimedia Communications Lab (KOM),\\
		Technische Universitaet Darmstadt, Germany}
}

\IEEEoverridecommandlockouts

\begin{document}
	\alglanguage{pseudocode}
\maketitle

\begin{abstract}
\input{Sections/0_abstract}

\end{abstract}
\IEEEpeerreviewmaketitle
%

\input{Sections/1_introduction}

\input{Sections/Related_Work}
\input{Sections/Model}

\input{Sections/3_FJ_Systems_with_Nonrenewal_Input}

\input{Sections/7_Correlated_ServiceTimes}

\input{Sections/EverythingModulated}


\input{Sections/Trace_evaluation.tex}

\input{Sections/Discussion.tex}

\input{Sections/Appendix}

\ifthenelse{\boolean{longVersion}}{\input{Sections/Appendix_B} }

\section*{Acknowledgement}
This work has been funded by the German Research Foundation~(DFG) as part of project C03 within the Collaborative Research Center~(CRC) 1053 -- MAKI. Computational facilities provided by the Lichtenberg-High Performance Computer at TU Darmstadt are also gratefully acknowledged. 
\bibliographystyle{IEEEtran}
\bibliography{main}

\end{document}

%% file: Sections/control-wasiur.tex
\usepackage[colorlinks=false, linkcolor=red]{hyperref}

\usepackage{graphicx}
\usepackage{subcaption}
\usepackage{amsmath}
\usepackage{cleveref}
\usepackage{amssymb}
\usepackage{mathtools}
\usepackage{algorithm}
\usepackage{algpseudocode}
\usepackage{algpascal}
\usepackage{color}

\usepackage{enumitem}

\usepackage{txfonts}
\usepackage{bm} 

\graphicspath{ {Figures/} }
\usepackage{subcaption}

\usepackage[T1]{fontenc}
\usepackage{cite}
\usepackage{dsfont}

\makeatletter
\algnewcommand\Input{\item[\textbf{Input:}]}%
\algnewcommand\Output{\item[\textbf{Output:}]}%
\algnewcommand\Init{\item[\textbf{Init:}]}%
\makeatother

\usepackage{amstext} 
\usepackage{array}   
\newcolumntype{L}{>{$}l<{$}} 
\newcolumntype{C}{>{$}c<{$}} 
\newcolumntype{R}{>{$}r<{$}} 

\usepackage{amsthm}
\theoremstyle{plain}
\newtheorem{myTheorem}{Theorem}

\theoremstyle{definition}
\newtheorem{myRemark}{Remark}
\newtheorem{myDefinition}{Definition}

\newcommand{\setOfReals}{\mathbb{R}}
\newcommand{\setOfNaturals}{\mathbb{N}}
\newcommand{\setOfNonnegativeIntegers}{\mathbb{N}_0}
\newcommand{\setOfPositiveReals}{\setOfReals_{+}}
\newcommand{\setN}[1]{ [ #1]}
\newcommand{\borel}[1]{\mathcal{B} (#1 )}
\newcommand{\effectiveDomain}[1]{\mathcal{D}(#1)}

\newcommand{\ExpDistribution}[1]{  \text{Exponential} (#1 ) }

\newcommand{\cardinality}[1]{\mid #1  \mid}

\DeclareMathOperator*{\esssup}{ess\,sup}

\newcommand{\indicator}[1]{\mathds{1}(#1)}

\newcommand{\myExp}[1]{\exp \bigl( #1 \bigr)  }

\newcommand{\E}{\mathrm{E}}
\newcommand{\Eof}[1]{\E[#1]}
\newcommand{\EofP}[2]{\E_{#1}\left(#2\right)}

\newcommand{\prob}{\mathrm{P}}
\newcommand{\probOf}[1]{\prob(#1)}

\newcommand{\eqpunkt}{.}
\newcommand{\eqkomma}{,}
\newcommand{\defeq}{\coloneqq}
\newcommand{\disteq}{\, =_{\text{D}} \, }

\newcommand{\ie}{\textit{i.e.}}
\newcommand{\eg}{\textit{e.g.}}

\usepackage{todonotes}
\usepackage{ifthen}

\newboolean{draftversion}
\setboolean{draftversion}{false} 

\newboolean{longVersion}
\setboolean{longVersion}{true}  

\newcommand{\proofLocation}[2]{\ifthenelse{\boolean{longVersion}}{\text{#1}}{\text{#2}}}

\newcommand{\tdWasiur}[2][]{\ifthenelse{\boolean{draftversion}}{\todo[inline, color=blue!20, caption={2do}, #1]{\begin{minipage}{\textwidth-4pt}\emph{Remark Wasiur:}\\#2\end{minipage}}}{}}

\newcommand{\tdAmr}[2][]{\ifthenelse{\boolean{draftversion}}{\todo[inline, color=orange!20, caption={2do}, #1]{\begin{minipage}{\textwidth-4pt}\emph{ToDo for Amr:}\\#2\end{minipage}}}{}}

\newcommand{\remarkAlex}[2][]{\ifthenelse{\boolean{draftversion}}{\todo[inline, color=red!20, caption={2do}, #1]{\begin{minipage}{\textwidth-4pt}\emph{Remark Alex:}\\#2\end{minipage}}}{}}

\newcommand{\tdGeneral}[2][]{\ifthenelse{\boolean{draftversion}}{\todo[inline, color=green!20, caption={2do}, #1]{\begin{minipage}{\textwidth-4pt}\emph{ToDO:}\\#2\end{minipage}}}{}}

\usepackage{url}

\usepackage{tabularx} 
\usepackage{xcolor}
\usepackage{xkeyval}
\input{tudcolours.def}

\usepackage{tikz}
\usetikzlibrary{positioning,fit,shapes,backgrounds,circuits.logic.US}

\tikzstyle{server}=[circle, line width=0.5pt, rounded corners=0.1mm, draw=black!100, fill=tud3a!100]
\tikzstyle{vertex}=[circle, line width=0.5pt, draw=black!100, fill=tud0a!50]
\tikzstyle{dispatcher} =[and gate US, line width=0.5pt, draw=black!100, fill=tud1a!100]
\tikzstyle{dotbox} = [draw=white, fill=white, rectangle,  inner sep=10pt, inner ysep=20pt]

\tikzset{three_sided/.style={
		draw=none,rectangle, 
		append after command={
			[shorten <= -0.5\pgflinewidth]
			([shift={(-1.5\pgflinewidth,-0.5\pgflinewidth)}]\tikzlastnode.north west)
			edge([shift={( 0.5\pgflinewidth,-0.5\pgflinewidth)}]\tikzlastnode.north east)
			([shift={( 0.5\pgflinewidth,-0.5\pgflinewidth)}]\tikzlastnode.north east)
			edge([shift={( 0.5\pgflinewidth,+0.5\pgflinewidth)}]\tikzlastnode.south east)
			([shift={( 0.5\pgflinewidth,+0.5\pgflinewidth)}]\tikzlastnode.south east)
			edge([shift={(-1.0\pgflinewidth,+0.5\pgflinewidth)}]\tikzlastnode.south west)
		}
	}
}

%% file: Sections/0_abstract.tex
Frameworks, such as MapReduce and  Hadoop are abundant nowadays. They seek to reap benefits of parallelization, albeit subject to a synchronization constraint at the output. Fork-Join (FJ) queuing models are used to  analyze such systems. Arriving jobs are split into tasks each of which is 
mapped to exactly one server. A job leaves the system when all of its tasks are executed. 

As a metric of performance, we consider waiting times for both work-conserving and non-work conserving server systems under a mathematical set-up general enough to take into account possible phase-type behavior of the servers, and as suggested by recent evidences,  bursty arrivals.

To this end, we present a Markov-additive process framework for an FJ  system and provide computable  bounds on  tail probabilities of  steady-state waiting times,  for both types of servers separately. We apply our results to three scenarios, namely, non-renewal (Markov-modulated) arrivals, servers showing phase-type behavior, and  Markov-modulated arrivals and services. We compare our  bounds against estimates obtained through simulations and  also provide a  theoretical conceptualization of  provisions 
in FJ systems. Finally, we calibrate our model with real data traces, and illustrate how our bounds can be used to devise  
provisions.

%% file: Sections/1_introduction.tex
\section{Introduction}
\label{sec:intro}
Recent infrastructural advancement of cloud computing and large-scale data processing has brought about massive deployment of parallel-server systems. 
Frameworks such as MapReduce \cite{dean2008mapreduce,polato2014comprehensive}, and its implementation Hadoop \cite{Hashem2016} are plentiful in today's world. Such systems seek to 
reap  the benefits of parallelization. However, often they are also subject to a synchronization constraint, because the final output is composed of outputs from all the servers. This makes performance evaluation of 
such systems  interesting.  Fork-Join (FJ) queuing models naturally capture the dynamics of system parallelization under synchronization constraint and  have been used to 
analyze such systems.

\begin{figure}
	\centering
	\includegraphics[width=1.0\linewidth]{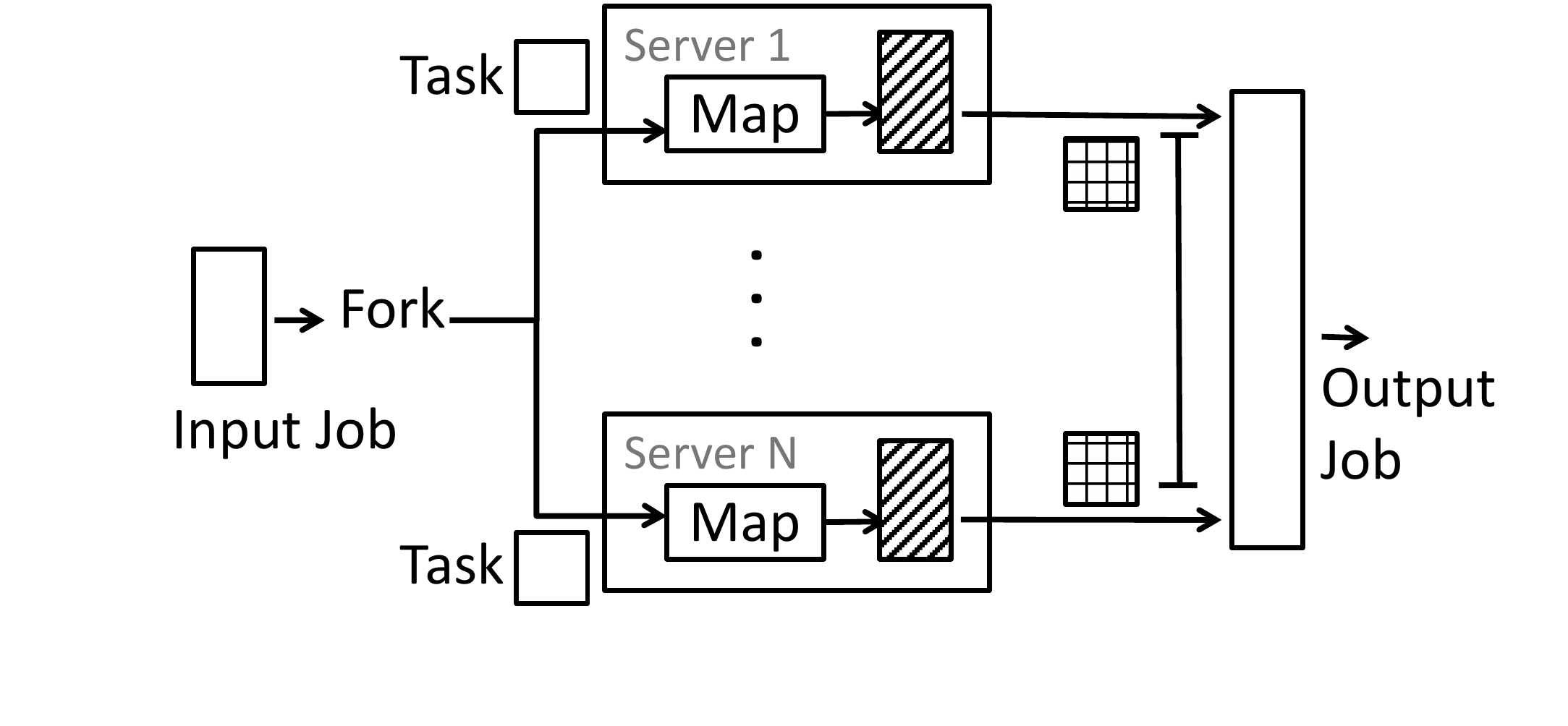}
	\caption{MapReduce as Fork-Join system.}
		\vspace{-1pt}
	\label{fig:mapReduce}
	\vspace{-2pt}
\end{figure}

In Fig.~\ref{fig:mapReduce} we present a MapReduce abstraction  resembling an FJ system.
Arriving jobs are first split into tasks each of which is subsequently mapped exactly to one 
 server executing the \emph{map} operation.
A job  leaves the system when all of its tasks are executed. We categorize the servers depending on whether they are work-conserving or not. Servers that start servicing the  task of the next job, if available, immediately after finishing the current job, are labeled  work-conserving. Servers that are not work-conserving, referred to as ``blocking" servers hereinafter, wait until \emph{all}  servers finish servicing their current jobs before starting the task of the next job. Blocking servers impose an additional synchronization barrier at the input. 

As a metric of performance in an FJ system, we consider 
the waiting time, which we define as the amount of time a job waits until its \emph{last} task starts being serviced. Its stochastic behavior is governed by the nature of  jobs arriving, \ie, the arrival process, and the service times of the servers. In the simplest case, one assumes a renewal process as arrival, and independent and identically distributed (iid) service times. However, recent evidences suggest that this 
assumption is untenable for various reasons. The arrival process may not be renewal and exhibit burstiness, \eg, input to a MapReduce, Internet traffic (see \cite{chen2012interactive,kandula2009nature,Yoshihara2001,Heffes1986MM}), and 
the servers may also be dependent in some sense. Therefore, we need a mathematical framework capable of accounting for these behaviors. In this paper, we present  a Markov additive process \cite{Iscoe1985largeDev} framework for this purpose (see Fig.~\ref{fig:graphical_model}) and show how particular application scenarios can be derived as special cases of it. In particular, we cover three  application scenarios: \textbf{(a)} non-renewal (Markov-modulated) arrivals, \textbf{(b)} servers showing phase-type behavior, and \textbf{(c)} Markov-modulated (MM) arrivals and services. 
We also bring in the notion of a
\emph{provision}, 
an umbrella term used  for a rule that decides job division into tasks, or that regulates service rates either \emph{reactively} or \emph{proactively}.     

An exact analysis of an FJ system with more than two servers in a general set-up remains elusive \cite{baccelli1989fork,boxma1994queueing}, because steady-state waiting time distribution is hard to obtain in closed form. 
One way to circumvent this problem is to attempt approximate results \cite{ko2008sojourn,nelson1988approximate,kemper2012mean,varma1994interpolation}.  Another approach, which we take in this paper, is to 
 bound the tail probabilities 
 \cite{Rizk2015Sigmetrics,Rizk2016}. 
 We also shed some light on how these bounds can be used for performance evaluation purposes of provisions. 
To give a concrete example, we also calibrate our model from a real data trace and devise a simplistic reactive provision. 


Our contributions in this paper are: \textbf{(1)} A Markov-additive process framework for a general FJ 
system and computable upper bounds on the tail probabilities of  steady-state waiting times, 
for work-conserving and blocking servers separately. \textbf{(2)} Application of our results to three scenarios, namely, non-renewal (Markov-modulated) arrivals, servers showing phase-type behavior, and  Markov-modulated arrivals and services. In the process, we also compare our theoretical bounds against empirical complementary cumulative distribution function (CCDF) obtained through simulations. \textbf{(3)} An abstract conceptualization of (reactive and proactive) provisions 
in FJ systems. \textbf{(4)} Calibration of our model from data traces 
and an example reactive provision for the purpose of illustration.

The paper is organized as follows: We highlight related work in Sec.~\ref{sec:related_work}. Sec.~\ref{sec:general_model} introduces the central mathematical model and presents the main results. In Sec.~\ref{sec:FJ_nonrenewal_input}, we apply our results to an FJ system with non-renewal input, followed by Sec.~\ref{sec:correlated_service_times} where we describe an FJ system with dependent servers and bring in the notion of provisions. 
An FJ system with Markov-modulated arrivals and services are discussed in Sec.~\ref{sec:everythingModulated}. Model calibration with real data traces is performed in Sec.~\ref{sec:trace_evaluation}.  We conclude the paper with a discussion in Sec.~\ref{sec:discussion}.

%% file: Sections/Related_Work.tex
\section{Related Work}
\label{sec:related_work}
The work most relevant to ours 
is \cite{Iscoe1985largeDev}, where the authors establish large deviations property of uniformly recurrent Markov additive processes. Later on several queuing theoretic results such as \cite{Duffield1994} have been derived based on \cite{Iscoe1985largeDev}.    Inequalities for the stationary waiting times  in  \emph{GI/G/k}  queues were first shown in \cite{kingman1970inequalities}. Martingale techniques have been used  to derive exponential upper bounds 
in \cite{buffet1994exponential, Duffield1994} and  in
\cite{poloczek2014scheduling}. 

Exact analysis of Fork-Join systems in a general set-up  is hard~\cite{baccelli1989fork,boxma1994queueing} 
and 
could only be 
carried out for a handful of special cases. Transient and  steady-state solutions of the FJ queue in terms of  virtual waiting times are obtained in \cite{Kim1989}.  Network calculus techniques  have  also been used to derive bounds \cite{Fidler2015infocom,fidler2016non}.
Results for 
 FJ systems with two servers having exponential service times under Poissonian job arrivals are shown in \cite{flatto1984two}. Useful approximations
\cite{ko2008sojourn,nelson1988approximate,kemper2012mean,varma1994interpolation}     and bounds  \cite{baccelli1989fork,balsamo1998bound,Rizk2015Sigmetrics,Rizk2016,Kesidis2015} have also surfaced.  

The authors in \cite{Xia2007} study the limiting behavior   of  FJ systems with blocking (finite buffer),  \ie, they study  the how the throughput of a general FJ system with blocking servers behaves as the 
 number of nodes increases to infinity while the processing speed and buffer space of each node remain  unaltered. In another interesting note,  FJ networks with non-exchangeable tasks under a heavy traffic (diffusion) regime  are studied in \cite{Atar2012}, where the authors show asymptotic equivalence between this network and its corresponding assembly network  with exchangeable tasks.

From the perspective of scheduling, \cite{harchol1999choosing} presents various policies  in a distributed server system and suggests  optimal ones for different situations. Similarly, \cite{Hyytia2012} attempts  to quantify the benefits parallelization in a dispatching system, where jobs, arriving in batches, are assigned to single-server FCFS
queues.
Note that the underlying premises in these works are quite dissimilar among themselves and from ours. 
The works \cite{baccelli1989fork,Rizk2016} are close to ours and share similar objective. 
While  \cite{Rizk2016} does consider Markov-modulated processes, they premise on homogeneous servers and only consider Markov-modulated arrivals on a state-space of size~$2$. On the other hand, \cite{baccelli1989fork} provides  bounds for   expected response times under renewal Poissonian arrival and exponential service times. 


Performance of 
MapReduce has been analyzed in \cite{dean2008mapreduce,zaharia2008improving}. The authors in  \cite{dean2008mapreduce} present  MapReduce as a programming model and show that many real world tasks are expressible in this model. On the other hand, \cite{zaharia2008improving} points out that  Hadoop’s performance
depends heavily  its task scheduler, which implicitly
assumes homogeneous cluster nodes, and
that it can be adversely impacted in a heterogeneous set-up.  To address this issue, 
they propose  
Longest Approximate Time
to End (LATE) scheduling algorithm.  Similar optimization problems 
are surveyed in \cite{polato2014comprehensive,Hashem2016}. In \cite{Lee2012parallel}, the authors discuss  pros and cons of MapReduce, and conclude efficiency issues, especially I/O costs    still need to be addressed. The efficiency of a MapReduce system, in general, requires tuning a number of parameters. In \cite{Babu2010automated}, Babu propose an  out-of-the-box automation technique to avoid manual tuning of the parameters. As opposed to our  theoretical standpoint, these articles provide a complimentary view from a practical implementation perspective. 
%
%

%% file: Sections/Model.tex
\section{The Model}
\label{sec:general_model}
In this section, we present our central mathematical model for Fork-Join systems. We  derive a general result from which we obtain several special cases that are relevant for practical purposes. In particular, we shall provide stochastic bounds on the steady state waiting  time distributions for a general heterogeneous setting. 

The following notational conventions are followed throughout the paper. We denote the set of natural numbers and the set of real numbers  by $\setOfNaturals$   and $\setOfReals$ respectively. Let  $\setOfNonnegativeIntegers \defeq \setOfNaturals \cup \{0\}$. For $N \in \setOfNaturals$, let $\setN{N} \defeq \{1,2,\ldots, N \}$. For $A \subseteq \setOfReals$, we denote the Borel $\sigma$-field of subsets of $A$ by $\borel{A}$. For any $f: \setOfReals \rightarrow \setOfReals$, we denote the effective domain of $f$ by $\effectiveDomain{f}$, \ie, $\effectiveDomain{f} \defeq \{ x \in \setOfReals \mid f(x) < \infty \}$. For an event $A$, we denote the indicator function of $A$ by $\indicator{A}$, taking value unity when $A$ is true and zero otherwise.



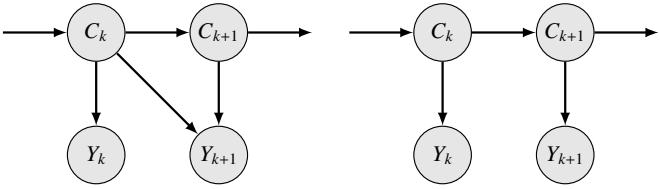
\begin{figure}
	\begin{subfigure}{0.48\columnwidth}
		\scalebox{0.85}{	\begin{tikzpicture} [inner sep=0cm, minimum size = 0.9cm]
			\node(vertex1) [vertex] {$C_k$} ;
			\node(vertex2) [vertex, right=1cm of vertex1] {$C_{k+1}$} ;
			\node(vertex3) [vertex, below=1cm  of vertex1] {$Y_k$} ;
			\node(vertex4) [vertex, below =1cm of vertex2] {$Y_{k+1}$} ;	
			\coordinate[right=1cm  of vertex2] (d1);
			\coordinate[left = 1cm of vertex1] (d2); 	
			\draw [-latex, line width=1pt,in=180, out=0] (vertex1) to node [minimum size=0.3cm, xshift=0.1cm, yshift=0.25cm] {} (vertex2);
			\draw [-latex, line width=1pt,in=90, out=-90] (vertex2) to node [minimum size=0.3cm, xshift=0.1cm, yshift=0.25cm] {} (vertex4);
			\draw [-latex, line width=1pt,in=135, out=-45] (vertex1) to node [minimum size=0.3cm, xshift=0.1cm, yshift=0.25cm] {} (vertex4);
			\draw [-latex, line width=1pt,in=90, out=-90] (vertex1) to node [minimum size=0.3cm, xshift=0.1cm, yshift=0.25cm] {} (vertex3);
			\draw [-latex, line width=1pt] (d2) to node [minimum size=0.3cm, xshift=0.1cm, yshift=0.25cm] {} (vertex1);
			\draw [-latex, line width=1pt] (vertex2) to node [minimum size=0.3cm, xshift=0.1cm, yshift=0.25cm] {} (d1);
			\end{tikzpicture} }
		\label{fig:Markov-additive}
	\end{subfigure}
	\hfill
	\begin{subfigure}{0.48\columnwidth}
		\scalebox{0.85}{	\begin{tikzpicture} [inner sep=0cm, minimum size = 0.9cm]
			\node(vertex1) [vertex] {$C_k$} ;
			\node(vertex2) [vertex, right=1cm of vertex1] {$C_{k+1}$} ;
			\node(vertex3) [vertex, below=1cm  of vertex1] {$Y_k$} ;
			\node(vertex4) [vertex, below =1cm of vertex2] {$Y_{k+1}$} ;	
			\coordinate[right=1cm  of vertex2] (d1);
			\coordinate[left = 1cm of vertex1] (d2); 	
			\draw [-latex, line width=1pt,in=180, out=0] (vertex1) to node [minimum size=0.3cm, xshift=0.1cm, yshift=0.25cm] {} (vertex2);
			\draw [-latex, line width=1pt,in=90, out=-90] (vertex2) to node [minimum size=0.3cm, xshift=0.1cm, yshift=0.25cm] {} (vertex4);
			\draw [-latex, line width=1pt,in=90, out=-90] (vertex1) to node [minimum size=0.3cm, xshift=0.1cm, yshift=0.25cm] {} (vertex3);
			\draw [-latex, line width=1pt] (d2) to node [minimum size=0.3cm, xshift=0.1cm, yshift=0.25cm] {} (vertex1);
			\draw [-latex, line width=1pt] (vertex2) to node [minimum size=0.3cm, xshift=0.1cm, yshift=0.25cm] {} (d1);
			\end{tikzpicture}}	
		\label{fig:Markov-modulated}
	\end{subfigure}
	\caption{Graphical representation of a Markov-additive process $\{C_k,Y_k  \}_{k \in \setOfNaturals} $ \textbf{(Left)} and  its special ``uncoupled'' case, the Markov modulated model \textbf{(Right)}. The nodes represent the variables and the arrows, the dependence structure. Please note that $Y_K$ is an additive component in that $Y_{k+1} = Y_k + Y_k^A$ for some $\{Y_k^A\}_{k \in \setOfNaturals}$. In  our case, $Y_k^A=  \sup_{n \in \setN{N}}     S_{n,k}  - A_{k} $. While the Markov-additive process, from the perspective of a provision, 
		is capable of modeling ``proactive'' systems (that anticipate the immediate future and act accordingly, \ie, set service rates accordingly) as well as reactive systems (that react on the current environment), the uncoupled model on the right is only capable of modeling the latter. 
	}
	\label{fig:graphical_model}
\end{figure}

\subsection{System description}
\label{sec:system_description}
Consider a single stage FJ queuing system with $N$ parallel servers as depicted in Fig.~\ref{fig:mapReduce}. 
Jobs arrive at the input station according to some  process with inter-arrival time $A_i$
between the $i$-th and $(i+1)$-th job, $i \in \setOfNaturals$. A job is  split into $N$ tasks each of which is  assigned to exactly one server.
The service time for the task of job $i$ at the $n$-th server is denoted by the random variable $S_{n,i}$, where $n \in \setN{N}$. To capture the effects of changing environment, we consider an underlying Markov chain $\{C_k\}_{k \in \setOfNonnegativeIntegers}$ on some general measure space $(\mathbb{E}, \mathcal{E})$. Note that $\mathbb{E}$ need not be finite, or even countable. We will explain later how different choices of $\mathbb{E}$  would fit different practical scenarios. We  need   two processes $\{ X_{n,k}\}_{n \in \setN{N}, k \in \setOfNonnegativeIntegers}$ and $\{Y_k\}_{k \in \setOfNonnegativeIntegers}$ defined  as follows
\begin{align}
X_{n,k} \defeq {} & \sum_{i=1}^{k} X_{n,i}^A \eqkomma   \;
Y_k \defeq {}  \sum_{i=1}^{k} Y_i^A\eqkomma
\label{eq:x_y_defn}
\end{align}
where $X_{n,i}^A= S_{n,i}  - A_{i}  $ and $Y_i^A= \sup_{n \in \setN{N}}     S_{n,i}  - A_{i}    $
for all $i \in \setOfNaturals$ and set  $X_{n,0}\defeq 0, Y_0 \defeq 0$, for each $n \in \setN{N}$.

\begin{figure*}[t!]
	\centering
	\begin{subfigure}[b]{0.32\textwidth}
		\centering
		\includegraphics[width=\textwidth]{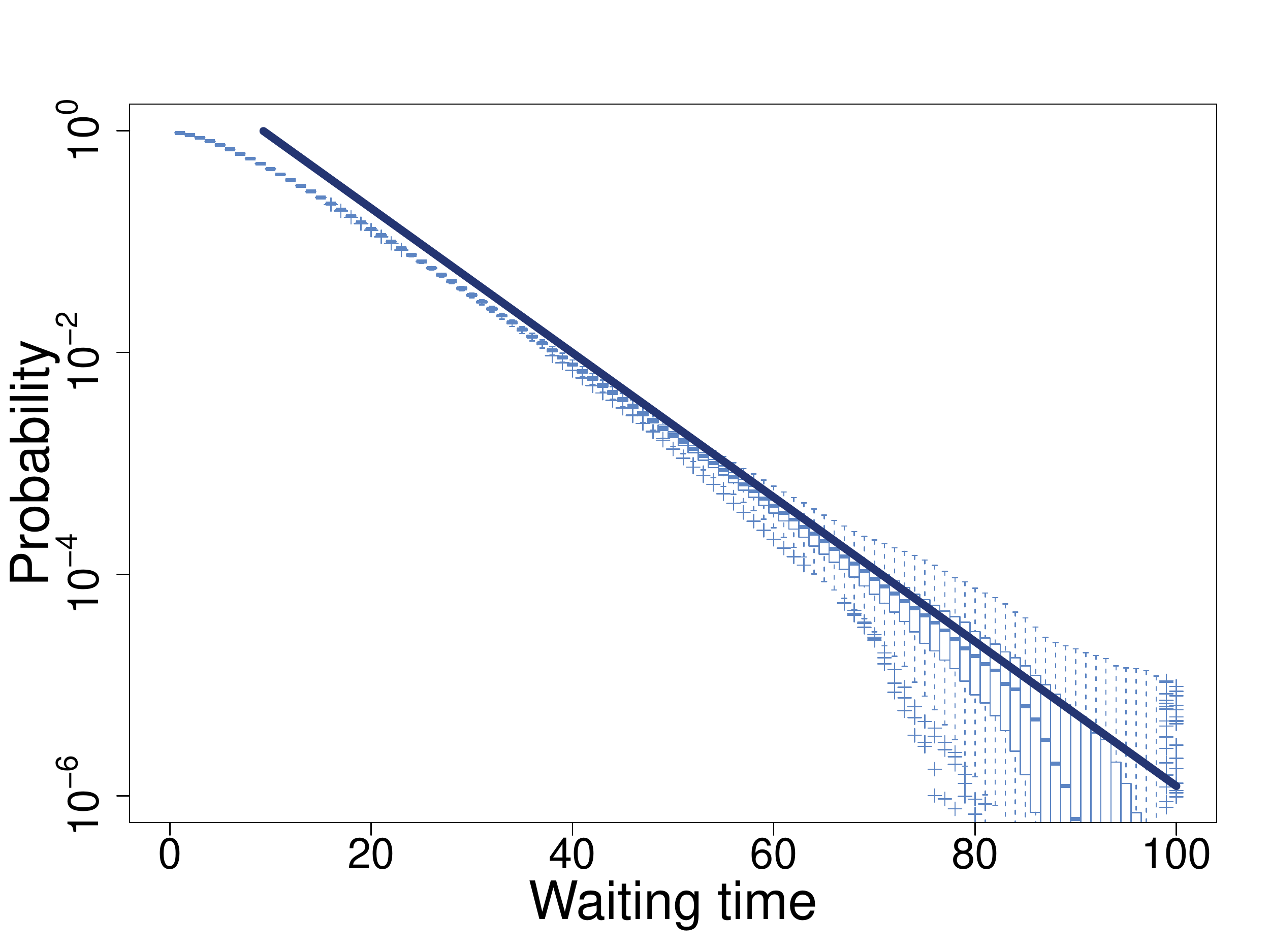}
	\end{subfigure}
	\begin{subfigure}[b]{0.32\textwidth}
		\centering
		\includegraphics[width=\textwidth]{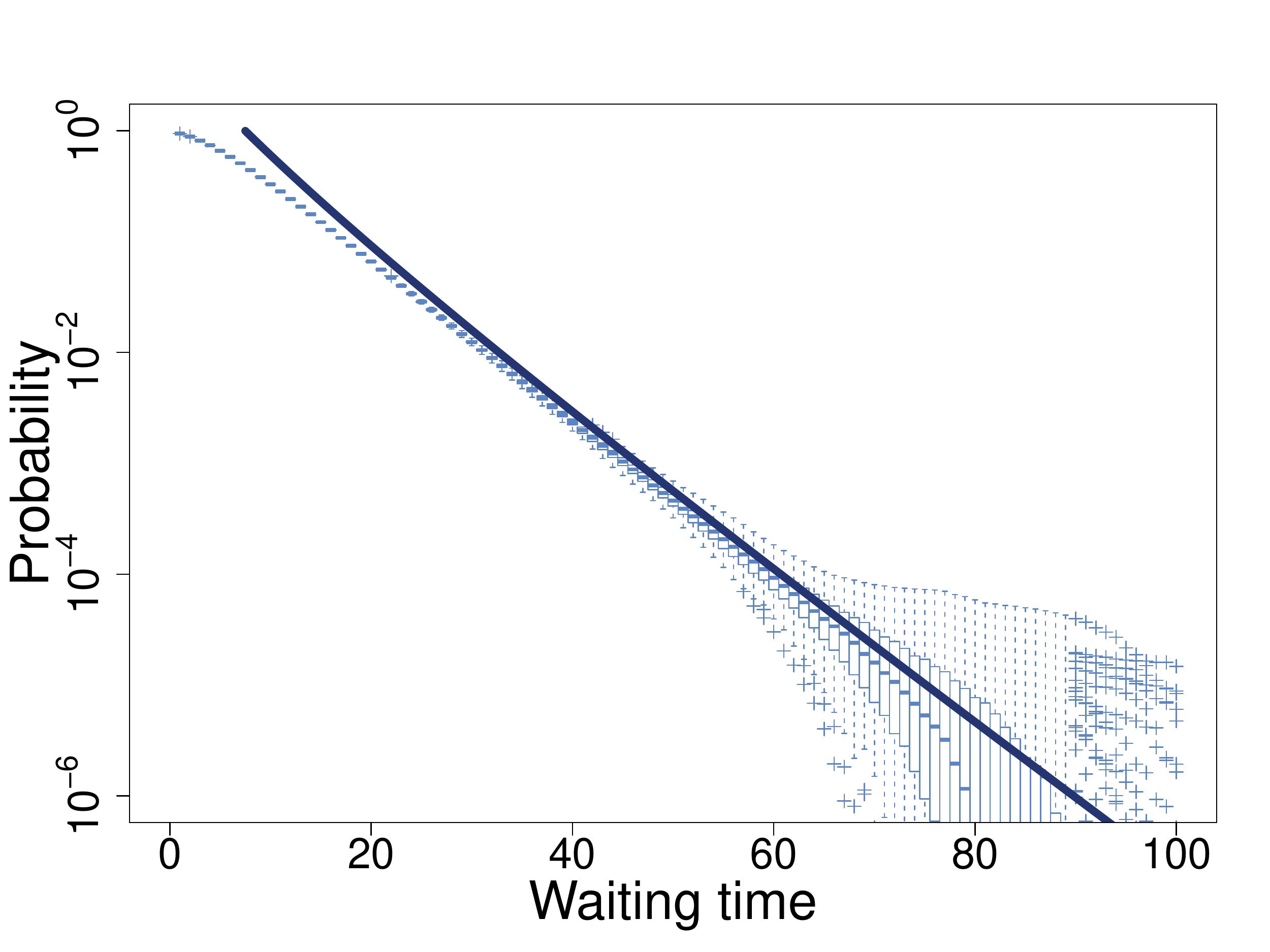}
	\end{subfigure}
	\begin{subfigure}[b]{0.32\textwidth}
		\centering
		\includegraphics[width=\textwidth]{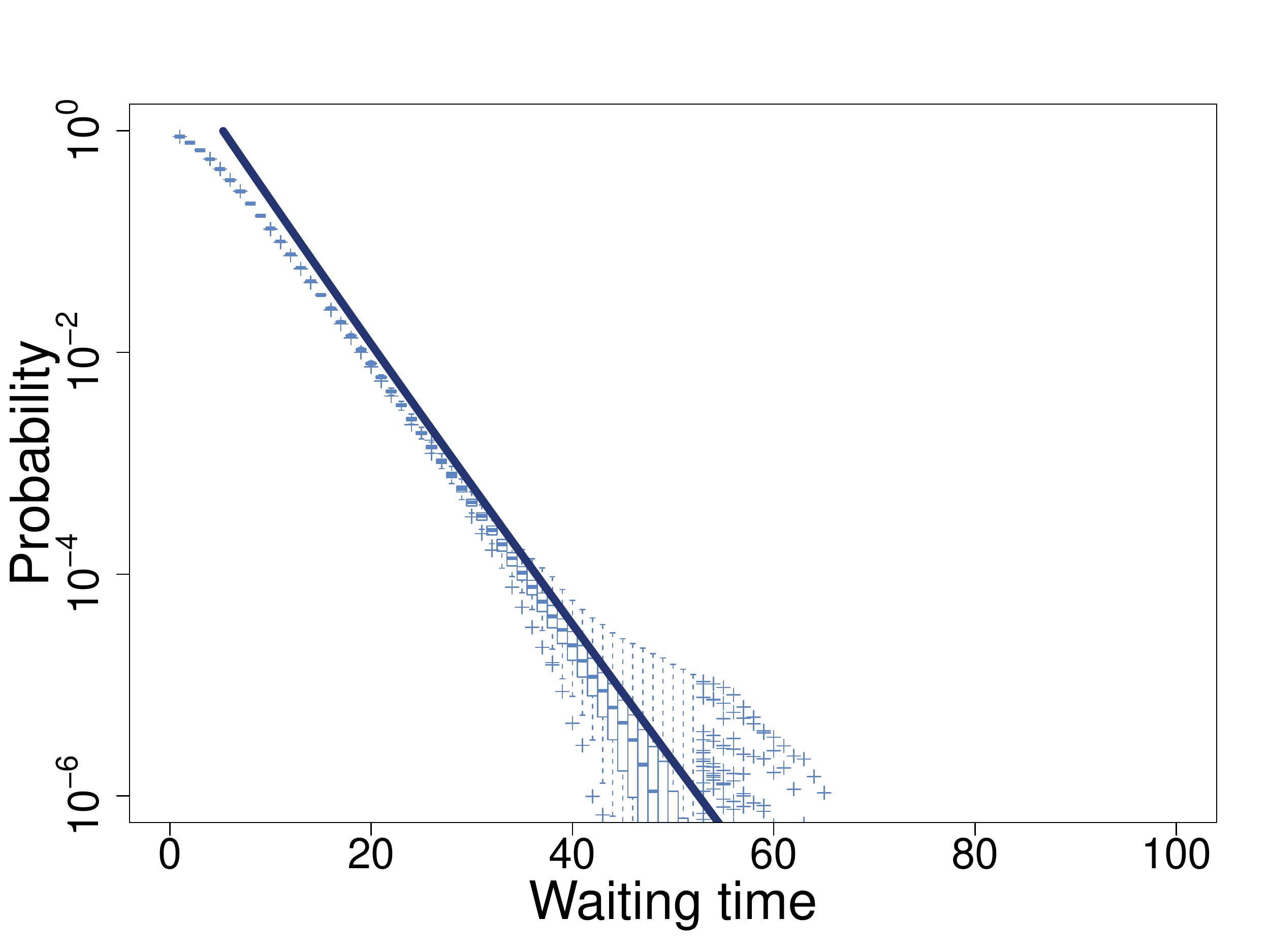}
	\end{subfigure}
	\caption{Numerical verification of the bounds (shown in darker shade) for work-conserving systems.  \textbf{(Left)} FJ system with Markov-modulated arrivals. The modulating Markov chain takes values in the set $\mathbb{E}= \{1,2,3,4\} $. 
		The exponential inter-arrival times have parameters $0.70, 0.75, 0.90$, and $0.95$. 
		\textbf{(Middle)} FJ system with Markov-modulated service times. The modulating Markov chain takes values in the set $\mathbb{E}= \{1,2,3,\ldots,32\} $. 
		The exponential inter-arrival times have parameter $ 0.9$. 
		\textbf{(Right)} FJ system with Markov-modulated arrival and service times. The modulating Markov chain takes values in the set $\mathbb{E}= \{1,2,3,\ldots,64\} $. 
		In all the cases, there are five heterogeneous and work-conserving servers whose service rates drawn randomly, satisfying the stability conditions in \ref{itm:as3} and \ref{itm:as4}.  The transition probabilities and the initial distribution of $C_k$ are chosen randomly.  }
	\label{fig:work_conserving_bounds}
\end{figure*}

We  assume that conditional on $\{C_k= c\}$, the servers act independently. In addition to this, we  assume that the processes $\{(C_k, X_{n,k})  \}_{k \in \setOfNaturals}$, for each $n \in \setN{N}$, and $\{ (C_k,Y_k ) \}_{k \in \setOfNaturals}$  are Markov-additive (MA) processes on $(\mathbb{E} \times \setOfReals, \mathcal{E}\times \borel{\setOfReals}   )$. Please refer to  Appendix~\ref{sec:AppendixA} for a precise definition and also see Fig.~\ref{fig:graphical_model}. 
Their transition kernels are defined as,  for  $ n \in \setN{N}$,
		\begin{equation}
		\begin{aligned}
	K_n(c,   A \times B  ) \defeq {} & \probOf{ (C_{1}, X_{n,1}) \in A \times B  \mid C_{0} = c }   \eqkomma \\
 \text{and }  L (c,   A \times B  ) \defeq {} & \probOf{ (C_{1}, Y_{1}) \in A \times B  \mid C_{0} = c } \eqkomma
	\end{aligned} \label{eq:kernel_defn}
		\end{equation}
		where   $ c \in \mathbb{E}\, ,   A \in \mathcal{E}$ and $B \in \borel{\setOfReals}$.
Also, to avoid triviality, 
we  assume a stable system with finite cumulants of the additive components. Define 
\begin{equation}
\begin{aligned}
\lambda^{(n)} (s)  \defeq {} &  \lim_{k \rightarrow \infty } k^{-1} \log \Eof{  \myExp{  s X_{n,k}  }  }  \eqkomma \\
\zeta (s)  \defeq {} &  \lim_{k \rightarrow \infty } k^{-1} \log \Eof{  \myExp{  s Y_{k}  }  }  \eqkomma
\end{aligned}
\end{equation}
allowing possibly infinite values. We make the technical assumptions precise in Appendix~\ref{sec:AppendixA}.

Having described our system, we look at one of the important metrics of performance, namely, waiting times.

\subsection{Waiting times}
\label{sec:waiting_time_defn}
\paragraph{Work-conserving servers}
For an FJ queuing system with $N$ work-conserving servers, we adopt the definition of waiting time $W_j$, for the $j$-th job, from \cite{Rizk2015Sigmetrics}, as 
$0$ for $j=1$ and $	\max\{ 0, \sup_{k \in \setN{j-1}} \{  \sup_{n \in \setN{N}} \{ \sum_{i=1}^{k} S_{n,j-i}  -\sum_{i=1}^{k} A_{j-i}  \} \}   \} \eqkomma $ for $j>1$. 
Intuitively we consider a job to be  waiting until its last task starts being serviced. Hence,  we have the following steady state representation of the waiting time $W$:
\begin{align}
W  \disteq  \sup_{k \in \setOfNonnegativeIntegers}  \sup_{n \in \setN{N}}   X_{n,k}  \eqkomma 
\label{eq:steady_state_nonblocking}
\end{align}
where $\disteq$ denotes equality in distribution and $X_{n,k} $ is as defined in \eqref{eq:x_y_defn}.

\paragraph{Blocking servers}
As before, for 
an FJ queuing system with $N$ blocking servers,   we adopt the definition of waiting time $W'_j$, for the $j$-th job, from \cite{Rizk2015Sigmetrics},   
as $0$ for $j=1$ and $	\max\{ 0, \sup_{k \in \setN{j-1}} \{  \sum_{i=1}^{k}  \sup_{n \in \setN{N}} S_{n,j-i}  -\sum_{i=1}^{k} A_{j-i}   \}   \} \eqkomma $ for $j>1$.  
Then,  we have the following steady state representation of the waiting time $W'$ in terms of $Y_k$ from \eqref{eq:x_y_defn}:
\begin{align}
W'  \disteq  \sup_{k \in \setOfNonnegativeIntegers}   Y_{k}   \eqpunkt
\label{eq:steady_state_blocking}
\end{align}

%

\subsection{Probabilistic bounds on waiting times}
In this section, we provide probabilistic bounds on the steady-state waiting times defined in \eqref{eq:steady_state_nonblocking}, and \eqref{eq:steady_state_blocking}. In order to do so, we need to transform the transition kernels defined in~\eqref{eq:kernel_defn} as follows, for all $c \in \mathbb{E}, \, A \in \mathcal{E}$,
\begin{equation}
\begin{aligned}
\tilde{K}_n(c, A ; s) \defeq {} & \int_{\setOfReals} K_n(c, A\times dx ) \myExp{s x }, \, \forall \, n \in \setN{N}    \eqkomma \\
\tilde{L} (c, A ; s) \defeq {} & \int_{\setOfReals} L(c, A\times dy ) \myExp{s y } \eqpunkt
\end{aligned} \label{eq:transformed_kernel}
\end{equation}
The highest eigenvalues of the transformed kernels in~\eqref{eq:transformed_kernel} play a pivotal role in constructing  martingales based on which the bounds would be derived. We now present 
  two theorems 
the first of which considers a work-conserving system. 

\begin{myTheorem}
(Work-conserving systems) Consider an FJ system with $N$ parallel work-conserving servers, as described in Secs.~\ref{sec:system_description} and \ref{sec:waiting_time_defn}. Then, we have 
\begin{enumerate}
	\item For all  $n \in \setN{N}$  and $s \in \effectiveDomain{\lambda^{(n)}}$,  $\myExp{ \lambda^{(n)}(s) }$ is the simple maximal eigenvalue of $\tilde{K}_n$ and the corresponding right eigenfunction $\{ r_n(c, s); \,  c \in \mathbb{E}  \}$ satisfying 
	\begin{equation*}
	\myExp{ \lambda^{(n)}(s) } r_n(c, s)  = \int_{\setOfReals} \tilde{K}_n(c, d\tau ; s) r_n(\tau, s) \eqkomma
	\end{equation*}
	is positive and bounded above.
	\item The tail probabilities of the steady-state waiting times defined in  \eqref{eq:steady_state_nonblocking}  are bounded above by 
	\begin{align}
	\probOf{W \geq w } \leq \sum_{n \in \setN{N}}  \phi_n( \theta_n ) \myExp{- \theta_n w} \eqkomma 
	\label{eq:main_nonblocking_bound}
	\end{align}
	where $\theta_n \defeq \sup \{ s>0 \mid  \lambda^{(n)}(s) \leq 0 \}$ and $ \phi_n (s) \defeq \esssup \{ \indicator{ X_{n,1} >0   } / r_n( C_1, s)   \}$, after having normalized $r_n(., \theta_n)$ so that 
	$\Eof{ r_n(C_0, \theta_n)   } =1$, for each $n \in \setN{N}$. 
\end{enumerate}
\label{thm:nonblocking_general}
\end{myTheorem}

The proof follows by extending already known results for Markov-additive processes from probability  literature (see \eg, \cite{Iscoe1985largeDev,Duffield1994}). However, for the sake of completeness, it is provided in  \ifthenelse{\boolean{longVersion}}{Appendix~\ref{sec:AppendixB}}{\cite{KhudaBukhsh2016techreport}}.  Now we provide bounds for the blocking system in the following theorem.

\begin{myTheorem}
(Blocking systems) Consider an FJ system with $N$ parallel work-conserving servers, as described in Secs.~\ref{sec:system_description} and \ref{sec:waiting_time_defn}. Then, we have 
\begin{enumerate}
	\item For all $s \in \effectiveDomain{\zeta}$,  $\myExp{\zeta(s) }$ is the simple maximal eigenvalue of $\tilde{L}$ and the corresponding right eigenfunction $\{ r(c , s); \,  c \in \mathbb{E}  \}$ satisfying 
	\begin{equation*}
	\myExp{ \zeta(s) } r(c, s)  = \int_{\setOfReals} \tilde{L}(c, d\tau ; s) r(\tau, s) \eqkomma
	\end{equation*}
	is positive and bounded above.
	\item The tail probabilities of the steady-state waiting times defined in \eqref{eq:steady_state_blocking} are bounded above by 
	\begin{align}
	\probOf{W' \geq w } \leq  \phi( \theta ) \myExp{- \theta w} \eqkomma 
	\label{eq:main_blocking_bound}
	\end{align}
	where $\theta \defeq \sup \{ s>0 \mid  \zeta(s) \leq 0 \}$ and $ \phi (s) \defeq \esssup \{ \indicator{ Y_{1} >0   } / r( C_1, s)   \}$ after having normalized $r(., \theta)$ so that 
	$\Eof{ r(C_0, \theta)   } =1$. 
\end{enumerate}
\label{thm:blocking_general}
\end{myTheorem}

The proof 
is provided in  \ifthenelse{\boolean{longVersion}}{Appendix~\ref{sec:AppendixB}}{\cite{KhudaBukhsh2016techreport}}. These two theorems are central to all the application scenarios that we consider in this paper. For ease of computation, we shall consider what is referred to as the ``uncoupled'' MA-process in \cite{Iscoe1985largeDev}. This essentially refers to a process with Markov-modulated increments (see Fig.~\ref{fig:graphical_model} and refer to \cite{Duffield1994}). 

\paragraph*{The ``uncoupled'' case} 
Suppose the distributions of increments,  $X_{n,k+1}^A$, for each $n \in \setN{N}$, and $ Y_{k+1}^A $ respectively for  work-conserving  and blocking systems, do not depend on $C_{k}$, conditional on $C_{k+1}$. This allows us to find conditional distributions $ Q_n(c, B) \defeq \probOf{X_{n,1}^A \in B \mid C_1=c}$, for each $n\in \setN{N}$ for  work-conserving systems, and $ R(c, B) \defeq \probOf{  Y_{1}^A \in B\mid C_1=c}$ for  blocking systems, for each $c  \in \mathbb{E}$ and $ B \in \borel{\setOfReals}$. Then, denoting the transition kernel of $\{ C_k \}$ alone by $T$, we simplify the transformed kernels in~\eqref{eq:transformed_kernel} as follows
\begin{equation*}
\begin{aligned}
\tilde{K}_n(c,   d\tau ; s )  =& T(c, d\tau) \int_{\setOfReals} Q_n(\tau, dz)  e^{ s z  }    
={} T(c, d\tau) \EofP{\tau}{ e^{ s X_{n,1}^A  }  } \eqkomma  \\
\tilde{L} (c, d\tau ; s) =& T(c, d\tau) \int_{\setOfReals} R(\tau, dz)  e^{ s z  } 
={}  T(c, d\tau) \EofP{\tau}{ e^{ s Y_{1}^A  }  }  \eqpunkt 
\end{aligned}
\end{equation*}
Here we use the shorthand notation $ \EofP{\tau}{ \myExp{ s X_{n,1}^A  }  }  $ to denote   $\Eof{ \myExp{ s X_{n,1}^A} \mid C_1=\tau }$, the moment generating function (MGF) of $X_{n,1}^A$ conditioned on $\{C_1=\tau\}$, the event that underlying Markov chain is in state~$\tau$. We can further simplify the formulas if we make following assumptions\footnote{These assumptions are only for the sake of simplification, the bounds hold  even without them.}. 
\begin{enumerate}[label=\color{tud0d}\textbf{A\arabic*}]
	\item \label{itm:b1} We assume that the service times and the arrival times are independent conditional on the $\{C_k=c\}$. This yields 
	\begin{equation} 
	\begin{aligned}
& 	\tilde{K}_n(c,   d\tau ; s ) = T(c, d\tau)   \EofP{\tau}{ e^{ s S_{n,1}  }  } \EofP{\tau}{ e^{ - s A_{1}  }  } \eqkomma \\
&	\tilde{L} (c, d\tau ; s)  = T(c, d\tau) \EofP{\tau}{ e^{ s \sup_{n \in \setN{N}}     S_{n,1}  }  } \EofP{\tau}{ e^{ - s A_{1}  }  }  \hspace{-2pt} \eqpunkt 
	\end{aligned} 
\label{eq:easy_transformed_kernel}
	\end{equation}
	\item \label{itm:b2} If further the increments $X_{n,1}^A$ and $Y_1^A$ take positive values with non-zero probability for any conditioning of $C_k$, then the essential supremums in Thms.~\ref{thm:nonblocking_general} and \ref{thm:blocking_general} simplify to 
	\begin{equation}
	\begin{aligned}
	\phi_n (s) = \sup_{c \in \mathbb{E}}  \{ 1 / r_n( c, s)   \} \eqkomma \; 
	\phi (s) = \sup_{c \in \mathbb{E}} \{ 1 / r( c, s)   \} \eqpunkt 
	\end{aligned} \label{eq:esssup_simplification}
	\end{equation}
\end{enumerate}
With these simplifications the computation of bounds on the tail probabilities of the waiting times is straightforward. We present the procedure in the form of  pseudocodes \ref{alg:non_blocking} and \ref{alg:blocking} for ease of understanding and implementation. 

\begin{algorithm}
	\caption{Pseudocode for work-conserving systems}
	\begin{algorithmic}[1]
		\Require Transition kernel $T$, and the MGFs $ \EofP{\tau}{ \myExp{ s S_{n,1} }  },  \EofP{\tau}{ \myExp{ - s A_{1}  }  } $
		\Ensure Stability \ref{itm:as3} and finiteness of cumulants \ref{itm:as4}
		\State Transform T to get $	\tilde{K}_n(c,   d\tau ; s ) $ 
		(see~\eqref{eq:easy_transformed_kernel})
		\For{$n \gets 1,\, N$} 
		\State $\myExp{  \lambda^{(n)}(s)} \gets $ maximal eigenvalue of $	\tilde{K}_n(c,   d\tau ; s ) $  
		\State $\theta_n \gets \sup \{ s>0 \mid  \lambda^{(n)}(s) \leq 0 \} $
		\State Normalize  $r_n(., \theta_n)$ so that 
		$\Eof{ r_n(C_0, \theta_n)   } =1$
		\State $	\phi_n (\theta_n ) \gets \sup_{c \in \mathbb{E}}  \{ 1 / r_n( c, \theta_n)   \} $
		\EndFor
		\State Compute bound on waiting time using~\eqref{eq:main_nonblocking_bound}
	\end{algorithmic} \label{alg:non_blocking}
\end{algorithm}

\begin{algorithm}
	\caption{Pseudocode for blocking systems}
	\begin{algorithmic}[1]
		\Require Transition kernel $T$, and the MGFs $ \EofP{\tau}{ \myExp{ s \sup_{n \in \setN{N}}     S_{n,i}  }  }, \EofP{\tau}{ \myExp{ - s A_{1}  }  } $
		\Ensure Stability \ref{itm:as3} and finiteness of cumulants \ref{itm:as4}
		\State Transform T to get $	\tilde{L} (c, d\tau ; s)  $ (see~\eqref{eq:easy_transformed_kernel})
		\State $\myExp{  \zeta(s)} \gets $ maximal eigenvalue of $	\tilde{L}(c,   d\tau ; s ) $ 
		\State $\theta \gets \sup \{ s>0 \mid  \zeta(s) \leq 0 \} $
		\State Normalize  $r(., \theta)$ so that 
		$\Eof{ r_n(C_0, \theta)   } =1$
		\State $	\phi (\theta ) \gets \sup_{c \in \mathbb{E}}  \{ 1 / r( c, \theta)   \} $
		\State Compute bound on waiting time using~\eqref{eq:main_blocking_bound}
	\end{algorithmic} \label{alg:blocking}
\end{algorithm}

Note that  both pseudocodes ~\ref{alg:non_blocking} and \ref{alg:blocking} require numerical solution methods when closed-form analytic expressions are difficult to obtain. Before closing the section, we make the following remark. 
\begin{myRemark}
The bounds in~\eqref{eq:main_nonblocking_bound} and \eqref{eq:main_blocking_bound} can also be used to derive bounds on the mean waiting times for the work-conserving and the blocking system respectively as follows
\begin{equation}
\begin{aligned}
\Eof{W} \leq {}  \sum_{n \in \setN{N}} \frac{ \phi_n(\theta_n) }{\theta_n} \; \text{ and }
\Eof{W'}  \leq {} \frac{ \phi(\theta)  }{\theta} \eqpunkt
\end{aligned} \label{eq:mean_bound}
\end{equation}
\end{myRemark}


 Next we apply our results to several scenarios in the following sections. They are intended to serve as simple illustrative examples. For ease of computation, assume that the state space $\mathbb{E}$ of the chain $\{C_k\}_{k \in \setOfNonnegativeIntegers}$ is finite. Then, the transition kernel~$T$ of $\{C_k\}_{k \in \setOfNonnegativeIntegers}$ is just a transition matrix. Let us write $T = (( t_{i,j}  ))$. We do allow the servers  to follow different probability distributions satisfying stability conditions \ref{itm:as3} and \ref{itm:as4}. We provide in the following simple examples where service and inter-arrival times are exponentially distributed. For the computation of MGF for the blocking system, we make use of  the following statistical result. 
 \begin{myRemark}
 	Consider a finite collection of independent random variables $\{U_n\}_{n \in \setN{N}}$ on $(\setOfPositiveReals^N, \borel{\setOfPositiveReals^N})$ such that  $U_n \sim \ExpDistribution{\mu_n}$  for each $n \in \setN{N}$. Write $\mu = (\mu_1, \mu_2, \ldots,\mu_n)$. Then, the mean and the moment generating function (MGF) of $V \defeq \max_{n \in \setN{N}} U_n$ are given by	
 	\begin{equation} \label{eq:alpha_beta_defn}
 	\begin{aligned}
  	\Eof{V} = &  	\alpha(\mu) \defeq \sum_{ \substack{ S \subset \setN{N}  \\ S \ne \emptyset } } (-1)^{\cardinality{S}+1 }  \frac{1  }{ (  \sum_{i \in S} \mu_i  ) } \eqkomma \\
 	\Eof{e^{s V } } = & 	\beta(\mu ;s ) \defeq 	 \sum_{ \substack{ S \subset \setN{N}  \\ S \ne \emptyset } } (-1)^{\cardinality{S}+1 } \frac{  (  \sum_{i \in S} \mu_i  )    }{(  \sum_{i \in S} \mu_i  )  -s} \eqpunkt
 	\end{aligned} 
 	\end{equation} 
 	\label[result]{result:max-of-exponential}
 \end{myRemark} 
 The proof is provided in  \ifthenelse{\boolean{longVersion}}{Appendix~\ref{sec:AppendixB}}{\cite{KhudaBukhsh2016techreport}}.

\begin{figure*}[t!]
	\centering
	\begin{subfigure}[b]{0.32\textwidth}
		\centering
		\includegraphics[width=\textwidth]{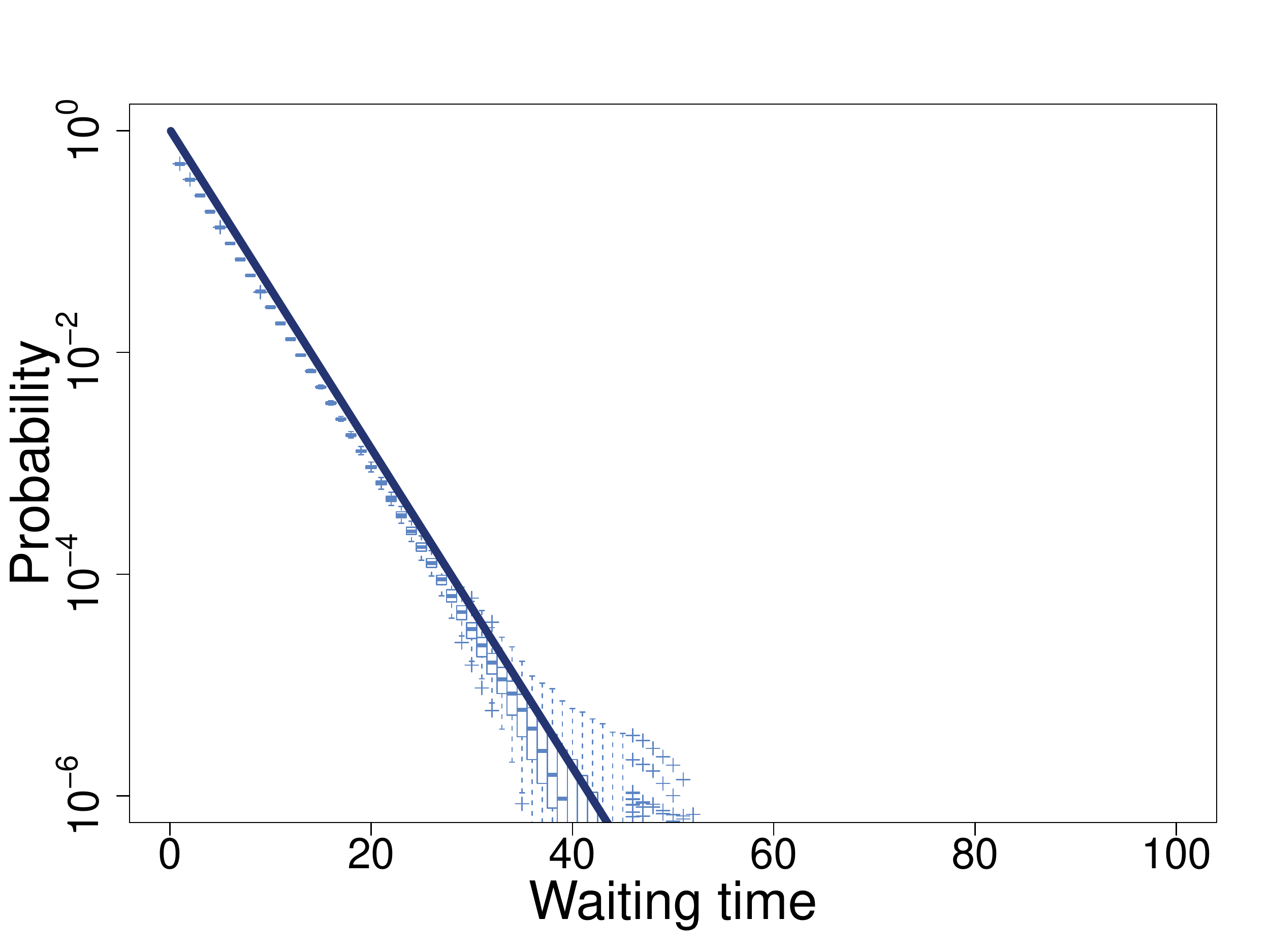}
	\end{subfigure}
	\begin{subfigure}[b]{0.32\textwidth}
		\centering
		\includegraphics[width=\textwidth]{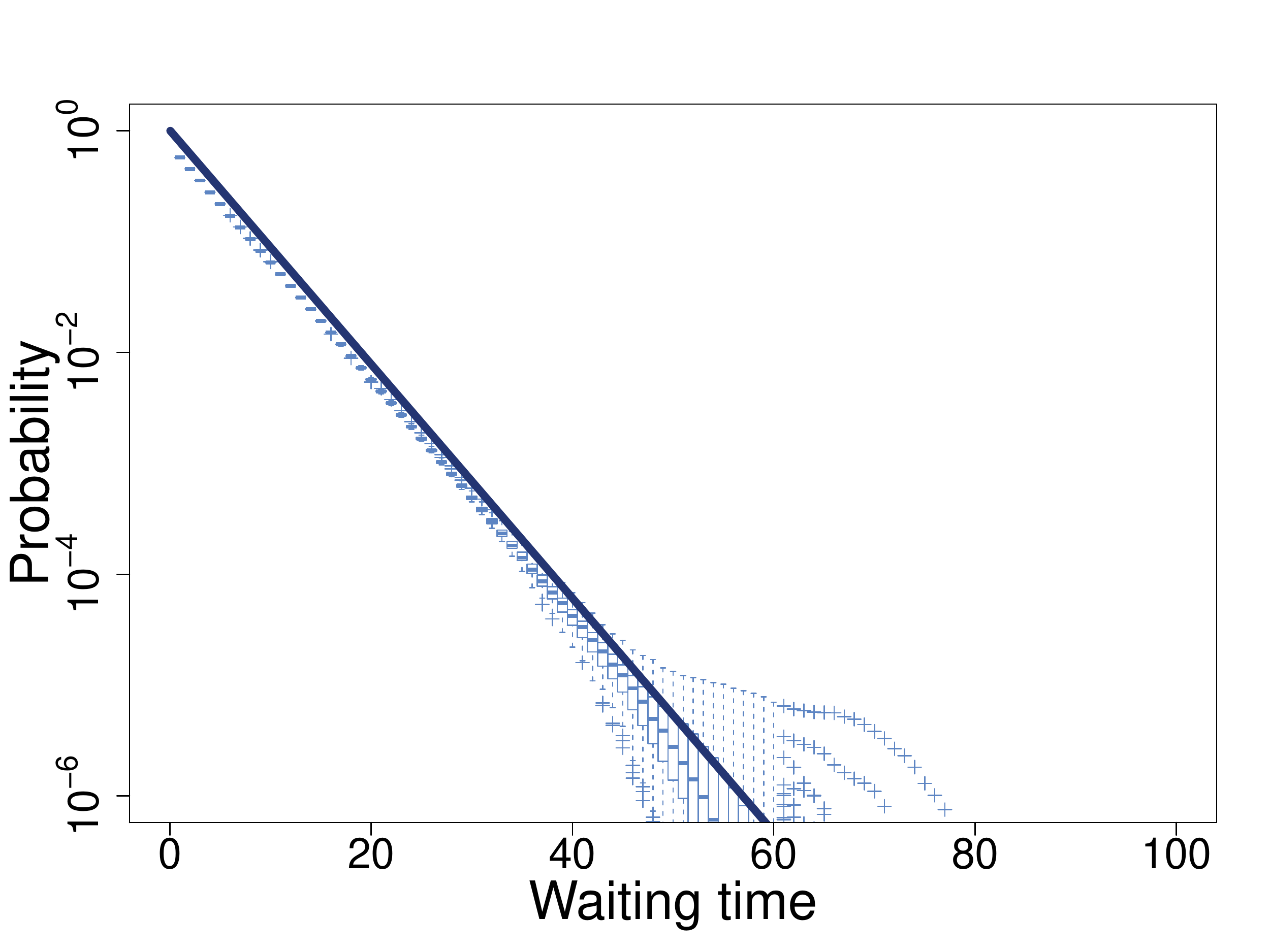}
	\end{subfigure}
	\begin{subfigure}[b]{0.32\textwidth}
		\centering
		\includegraphics[width=\textwidth]{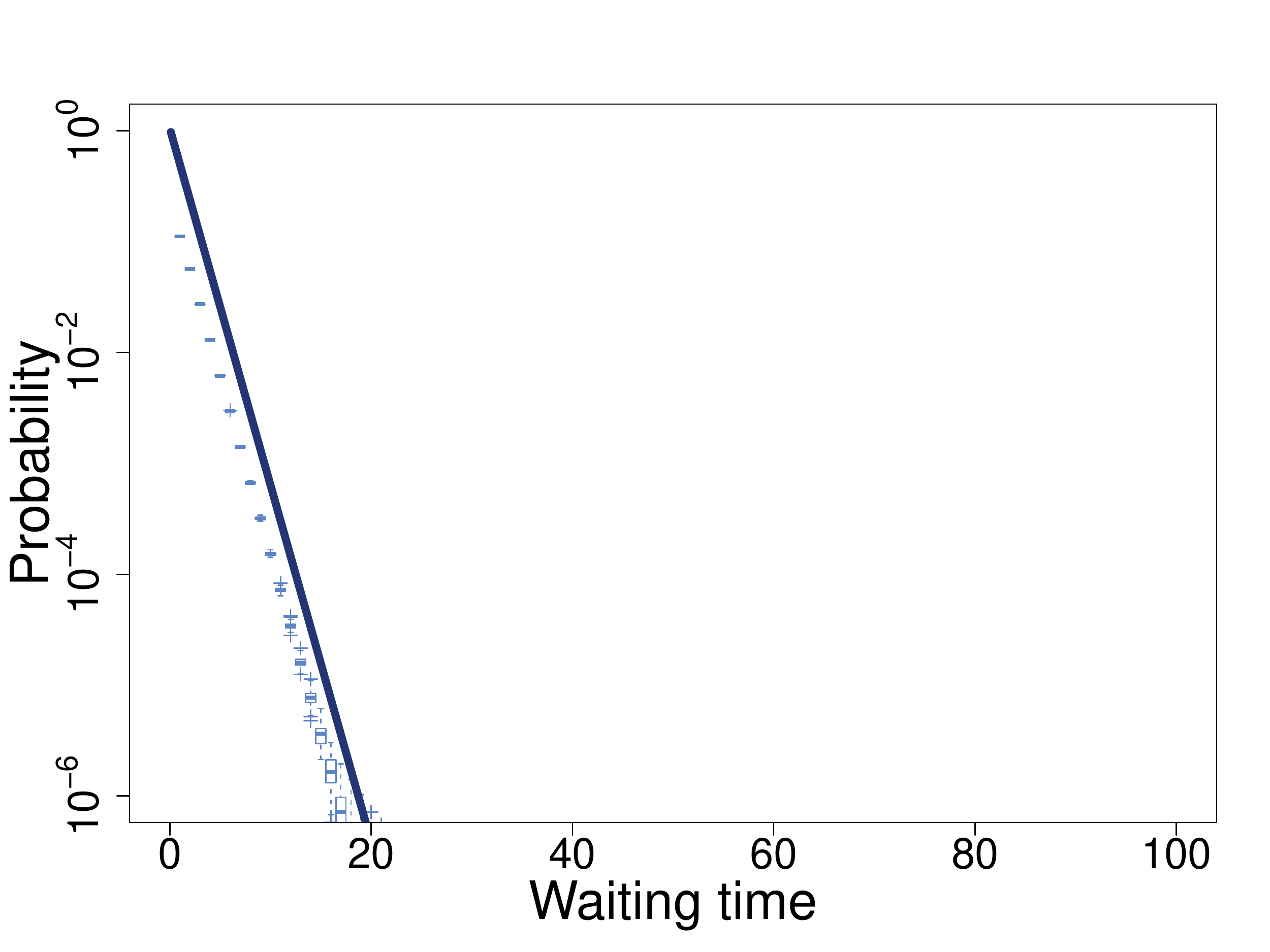}
	\end{subfigure}
	\caption{Numerical verification of the bounds (shown in darker shade) for blocking systems.  \textbf{(Left)} FJ system with Markov-modulated arrivals. The modulating Markov chain takes values in the set $\mathbb{E}= \{1,2,3\} $. 
		The exponential inter-arrival times have parameters $0.25, 0.4$, and $0.50$. 
		\textbf{(Middle)} FJ system with Markov-modulated service times. The modulating Markov chain takes values in the set $\mathbb{E}= \{1,2,3,\ldots,32\} $. 
		The exponential inter-arrival times have parameter $ 0.35$. 
		\textbf{(Right)} FJ system with Markov-modulated arrival and service times. The modulating Markov chain takes values in the set $\mathbb{E}= \{1,2,3,\ldots,64\} $. 
		In all the cases, there are five heterogeneous and blocking servers whose service rates drawn randomly, satisfying the stability conditions in \ref{itm:as3} and \ref{itm:as4}.  The transition probabilities and the initial distribution of $C_k$ are chosen randomly.  }
	\label{fig:blocking_bounds}
\end{figure*}

%% file: Sections/3_FJ_Systems_with_Nonrenewal_Input.tex
\section{FJ System with non-renewal input}
\label{sec:FJ_nonrenewal_input}
In this section, we describe an FJ system with Markov-modulated inputs. This is principally motivated by recent empirical evidences that reveal 
burstiness in Internet traffic  and also in inputs to MapReduce clusters \cite{chen2012interactive,kandula2009nature,Yoshihara2001,Heffes1986MM}. In general, to model this 
burstiness,  we can assume the inter-arrival times to be  modulated by some 
Markov chain $\{C_k\}_{k \in \setOfNonnegativeIntegers}$.

\subsection*{Numerical example: MM inter-arrival times} 
Suppose the modulating Markov chain takes four distinct values (corresponding to  different phases of arrival traffic). In state~$j$ of the chain, suppose the inter-arrival  times are 
exponentially distributed with parameter~$\lambda_j$. Also assume, the service times at the $n$-th server are exponentially distributed with parameter~$\mu_n$.  Then, the transformation in~\eqref{eq:easy_transformed_kernel} 
is simply 
\begin{equation*}
\left(\begin{matrix}
t_{11} & t_{12} & t_{13} & t_{14} \\
t_{21} & t_{22} & t_{23} & t_{24}\\
t_{31} & t_{32} & t_{33} & t_{34}\\
t_{41} & t_{42} & t_{43} & t_{44} \\
\end{matrix} \right)   
\rightarrow 
\left(\begin{matrix}
t_{11}  \frac{ \lambda_1  }{\lambda_1 + s}  & t_{12} \frac{ \lambda_2  }{\lambda_2 + s}  & t_{13}\frac{ \lambda_3  }{\lambda_3 + s}  & t_{14}\frac{ \lambda_4  }{\lambda_4 + s}  \\
t_{21}\frac{ \lambda_1  }{\lambda_1 + s}  & t_{22}\frac{ \lambda_2  }{\lambda_2 + s}  & t_{23}\frac{ \lambda_3  }{\lambda_3 + s}  & t_{24}\frac{ \lambda_4  }{\lambda_4 + s}  \\
t_{31} \frac{ \lambda_1  }{\lambda_1 + s} & t_{32} \frac{ \lambda_2  }{\lambda_2 + s} & t_{33} \frac{ \lambda_3  }{\lambda_3 + s} & t_{34}\frac{ \lambda_4  }{\lambda_4 + s}  \\
t_{41} \frac{ \lambda_1  }{\lambda_1 + s}  & t_{42} \frac{ \lambda_2  }{\lambda_2 + s} & t_{43} \frac{ \lambda_3  }{\lambda_3 + s} & t_{44}\frac{ \lambda_4  }{\lambda_4 + s}  \\
\end{matrix} \right)   \eqpunkt
\end{equation*}
Having done the above transformation, the decay rates are found as 
\begin{equation}
\begin{aligned}
\theta_n =  \sup \{ s>0 \mid   \frac{\mu_n}{\mu_n -s}  \chi_A(s) \leq 1   \}  \eqkomma \\
\theta =  \sup \{ s>0 \mid   \beta(\mu;s)  \chi_A(s) \leq 1   \}  \eqkomma
\end{aligned}
\end{equation}
where $\chi_A$ is the largest eigenvalue of the transformed matrix. After normalization of the right eigenvector, one obtains the bounds using~\eqref{eq:main_nonblocking_bound} and~\eqref{eq:main_blocking_bound} for the work-conserving  and the  blocking system respectively.  Please see Fig.~\ref{fig:work_conserving_bounds} (for work-conserving systems) and  Fig.~\ref{fig:blocking_bounds} (for blocking systems)  to compare  our bounds against  complementary cumulative distribution functions (CCDFs) obtained  from   simulations.

%% file: Sections/7_Correlated_ServiceTimes.tex
\section{Parallel Systems with Dependent Servers}
\label{sec:correlated_service_times}



In this section, we consider an FJ system as described in Sec.~\ref{sec:general_model} with correlated servers. To be precise, we assume that the service times are modulated by some  Markov chain. The motivation behind this is the phase type behaviors that service times show due to various external effects. Before furnishing numerical examples, we mention some  factors that might engender such a phase-type behavior. 
\paragraph{Shared resources} As depicted in Fig.~\ref{fig:shared_resources}, the servers may be shared resources and hence could only be partially utilized. 
The modulating chain 
allows us to model the share of server capacity utilizable  by each incoming job.  
\paragraph{Unequal job sizes} Many a time we are faced with situations where the job arrival process is  renewal, but the job sizes are unequal. In the context of MapReduce, the job sizes  could be time varying. In this case, the modulating chain would stand for different job sizes enforcing different service time distributions. The state space of the chain~$\mathbb{E}$ can be chosen depending on the particular application under consideration.  
\paragraph{Provisions  in MapReduce} The ``irregular'' service times may also arise due to 
 provisions, 
 even when the job sizes do not change. Suppose  that the incoming jobs are split unequally among the available servers. The rule that decides job division into tasks is termed a \emph{provision}.  
 Such provisions can be employed in MapReduce systems to manipulate waiting times. Consider a simple example. Each job consists of two sub-jobs one of which is more demanding than the other. That is, $\text{Job}_i = (\text{Job}_{i,1} ,\text{Job}_{i,2} )$, where $\text{Job}_{i,1}  $ can be assumed to be heavier without loss of generality. Now, in order to apportion the burden of the heavier job, devise a variant of round robin  mechanism such that for the first job $\text{Job}_1$, the sub-job $ \text{Job}_{1,1} $ is allotted to servers~$1,2,\ldots, d$ and $ \text{Job}_{1,2} $, to the rest. Then,  $ \text{Job}_{2,1} $ is allotted to servers~$d+1,d+2,\ldots, 2d$ and $ \text{Job}_{2,2} $ to the rest, and so on. Mathematically this is equivalent to having a modulating Markov chain that starts at state~$1$ where it assigns service rates appropriate of the heavier job to servers~$1,2,\ldots, d$ and the usual, to the rest, and then jumps with probability one to state~$2$  where  it assigns service rates appropriate of the heavier job to servers~$d+1,d+2,\ldots, 2d$ and the usual, to the rest. 
\paragraph{Modulation in MPTCP} Packet scheduling or load-balancing mechanisms could also give rise to correlated service times. The load-balancing algorithm
typically decides the amount of packets to send over each path with the objective of keeping  congestion under control, or to a minimum. Incorporating all the subtleties of a real system into a mathematical model is often not feasible. However, taking the liberty of mathematical abstraction, we can model such a scenario with a Markov chain (representing the decisions of the load-balancer) that modulates only the service times of the system. 
\paragraph{Efficiency differentiation} Servers may themselves have their  own high and low efficiency periods that may or may not depend on other servers (possibly enforced by energy-saving routines). 



\subsection*{Numerical example: MM service times} 
Motivated by the above scenarios, we now elaborate the bound computation. 
In the following example, assume the arrival process is renewal and inter-arrival times are exponentially distributed with parameter~$\lambda$.

Suppose there are two servers each of which has two efficiency phases, high and low. We can model this by two Markov chains modulating the servers, each on state space $\{0,1\}$.  For the sake of simplicity, assume that server~$i$ is exponentially distributed with parameter~$\mu_{i}$ or $\kappa_{i}$ according as its modulating Markov chain is state~$0$ or $1$.  The two Markov chains may or may not be independent. Mathematically this is equivalent to having one single modulating Markov chain on state space~$\{0,1\} \times \{0,1\}$. Since the set $\{0,1\} \times \{0,1\}$ has one-to-one correspondence with the set~$\{1,2,3,4\}$, we can conveniently rename the states as 
$(0,0) \rightarrow 1  \eqkomma  \, (0,1) \rightarrow 2  \eqkomma  \, (1,0) \rightarrow 3  \eqkomma  \, (1,1) \rightarrow 4$.
Now  let us first look at the work-conserving system. For the $1$st server, we  transform 
\begin{equation*}
\left(\begin{matrix}
t_{11} & t_{12} & t_{13} & t_{14} \\
t_{21} & t_{22} & t_{23} & t_{24}\\
t_{31} & t_{32} & t_{33} & t_{34}\\
t_{41} & t_{42} & t_{43} & t_{44} \\
\end{matrix} \right)   
\rightarrow 
\left(\begin{matrix}
t_{11}  \frac{ \mu_{1}   }{\mu_{1} - s}  & t_{12}  \frac{ \mu_{1}   }{\mu_{1} - s}  & t_{13} \frac{ \kappa_1   }{\kappa_1 - s}    & t_{14} \frac{ \kappa_1  }{\kappa_1 - s}   \\
t_{21}\frac{ \mu_{1}   }{\mu_{1}  - s}  & t_{22} \frac{ \mu_{1}   }{\mu_{1} - s}    & t_{23} \frac{ \kappa_1  }{\kappa_1 - s}  & t_{24} \frac{ \kappa_1  }{\kappa_1 - s}   \\
t_{31} \frac{ \mu_{1}   }{\mu_{1}  - s} & t_{32}  \frac{ \mu_{1}   }{\mu_{1} - s}   & t_{33} \frac{ \kappa_1   }{\kappa_1 - s}  & t_{34} \frac{ \kappa_1  }{\kappa_1- s}  \\
t_{41} \frac{ \mu_{1}   }{\mu_{1}  - s}  & t_{42} \frac{ \mu_{1}   }{\mu_{1} - s}   & t_{43}  \frac{ \kappa_1  }{\kappa_1 - s}   & t_{44} \frac{ \kappa_1 }{\kappa_1 - s}   \\
\end{matrix} \right)   \eqpunkt
\end{equation*}
Transformation for the $2$nd server is analogous. Denote the largest eigenvalues of these two transformed matrices  by $\chi_S^{(1)}$ and $\chi_S^{(2)}$ respectively. The transformation for the blocking system is as follows
\scalebox{0.95}{  \parbox{\linewidth}{%
	\begin{equation*}
	\left(\begin{matrix}
	t_{11} \beta(\mu_1, \kappa_1;s )  & t_{12}   \beta(\mu_1, \kappa_2;s) & t_{13}  \beta(\mu_2, \kappa_1;s)    & t_{14}  \beta(\mu_2, \kappa_2;s)    \\
	t_{21} \beta(\mu_1, \kappa_1;s)   & t_{22}  \beta(\mu_1, \kappa_2;s)    & t_{23}  \beta(\mu_2, \kappa_1;s)  & t_{24} \beta(\mu_2, \kappa_2;s) \\
	t_{31} \beta(\mu_1, \kappa_1;s)  & t_{32}   \beta(\mu_1, \kappa_2;s) & t_{33} \beta(\mu_2, \kappa_1; s)  & t_{34}  \beta(\mu_2, \kappa_2;s)  \\
	t_{41}  \beta(\mu_1, \kappa_1;s)  & t_{42}  \beta(\mu_1, \kappa_2;s)    & t_{43}  \beta(\mu_2, \kappa_1;s)  & t_{44} \beta(\mu_2, \kappa_2;s)   \\
	\end{matrix} \right)   \eqpunkt
	\end{equation*}    }}
Call its largest eigenvalue $\chi_{S}$. The function~$\beta$ is as defined in~\eqref{eq:alpha_beta_defn}. 
Having done the above transformation, the decay rates are found as 
\begin{equation}
\begin{aligned}
\theta_n =  \sup \{ s>0 \mid   \frac{\lambda}{\lambda +s}  \chi_S^{(n)}(s) \leq 1   \}  \eqkomma \\
\theta =  \sup \{ s>0 \mid   \frac{\lambda}{\lambda +s}   \chi_S(s) \leq 1   \}  \eqpunkt
\end{aligned}
\end{equation}
After normalization of the right eigenvector, one finds the bounds using formulas in~\eqref{eq:main_nonblocking_bound} and~\eqref{eq:main_blocking_bound} for the work-conserving  and the  blocking system respectively.  To see the quality of our bounds on a bigger state space, we simulated an FJ system with five heterogeneous servers being modulated by a chain having $32$ states. See Fig.~\ref{fig:work_conserving_bounds} (for work-conserving systems) and  Fig.~\ref{fig:blocking_bounds} (for blocking systems)  to compare  our bounds against  empirical CCDFs.

%% file: Sections/EverythingModulated.tex
\section{Markov Modulated Arrival and Services}
\label{sec:everythingModulated}
In this section, we describe a system where   service  and  inter-arrival times may be dependent. This is essentially a generalization over Secs.~\ref{sec:FJ_nonrenewal_input} and \ref{sec:correlated_service_times}. All the motivating examples listed in Secs.~\ref{sec:FJ_nonrenewal_input} and   \ref{sec:correlated_service_times} can be extended to this case to account for 
generalized application scenarios. While this allows us to endow service times of each server, and the arrival process, separate modulating Markov chains (which can be modeled by one single chain on the Cartesian product space as shown before), we can use this formalism to devise more advanced provisions 
too by 
taking into account the current job arrival rate (\ie, set efficiency of servers to ``high'' during busy period and to ``low'' otherwise etc.). This paves way for what we call ``reactive 
provisions.''
\paragraph{Reactive provisions} Recall the definition of a 
provision 
as discussed in Sec.~\ref{sec:correlated_service_times}. We propose to take into account information on the current environment in some form and then modulate 
(\ie, set service rates accordingly). Such a provision 
is reactive in nature and hence the nomenclature. The changing environment is essentially captured through the modulating Markov chain in this case.  
\subsection*{Numerical example: MM inter-arrival and service times}
Consider a Markov chain $(C_k)_{k \in \setOfNonnegativeIntegers}$ capturing the changing environment in the sense that at state~$j$ of the inter-arrival times are exponentially distributed with parameter~$\lambda_j$ and accordingly, the service times at the $n$-th server are distributed exponentially with parameter~$\mu_{n,j}$. Define $\mu^{(j)} \defeq (\mu_{1,j}, \mu_{2,j}, \ldots, \mu_{n,j})$. Then, the required transformation for  work-conserving systems is 
$t_{ij} \rightarrow t_{ij} \left( \frac{ \mu_{n,j}   }{\mu_{n,j} - s}\right)  \left( \frac{  \lambda_j  }{\lambda_j +s} \right) \eqkomma $
for the $n$-th server, and likewise, the transformation for the blocking system is given by 
$t_{ij} \rightarrow t_{ij} \beta(\mu^{(j)} ; s)  \left( \frac{  \lambda_j  }{\lambda_j +s} \right)  \eqpunkt$
Let us denote the largest eigenvalue of the transformed matrix for the $n$-th server by $\chi_{AS}^{(n)}$, and that of the transformed matrix for the blocking system by $\chi_{AS}$. Therefore, the decay rates are found as 
\begin{equation}
\begin{aligned}
\theta_n =  \sup \{ s>0 \mid   \chi_{AS}^{(n)}(s) \leq 1   \}  \eqkomma \\
\theta =  \sup \{ s>0 \mid     \chi_{AS}(s) \leq 1   \}  \eqpunkt
\end{aligned}
\end{equation}
After normalization of the right eigenvector, we compute the  bounds using formulas in~\eqref{eq:main_nonblocking_bound} and~\eqref{eq:main_blocking_bound}. 
To see the quality of our bounds, we simulated the system with the modulating chain having $64$ states. See Fig.~\ref{fig:work_conserving_bounds} for work-conserving systems and  Fig.~\ref{fig:blocking_bounds} for blocking systems  to compare  our bounds against empirical CCDFs.

%% file: Sections/Trace_evaluation.tex
\section{Trace-based Evaluation}
\label{sec:trace_evaluation}
\begin{figure}
	\centering
	\includegraphics[width=0.35\textwidth]{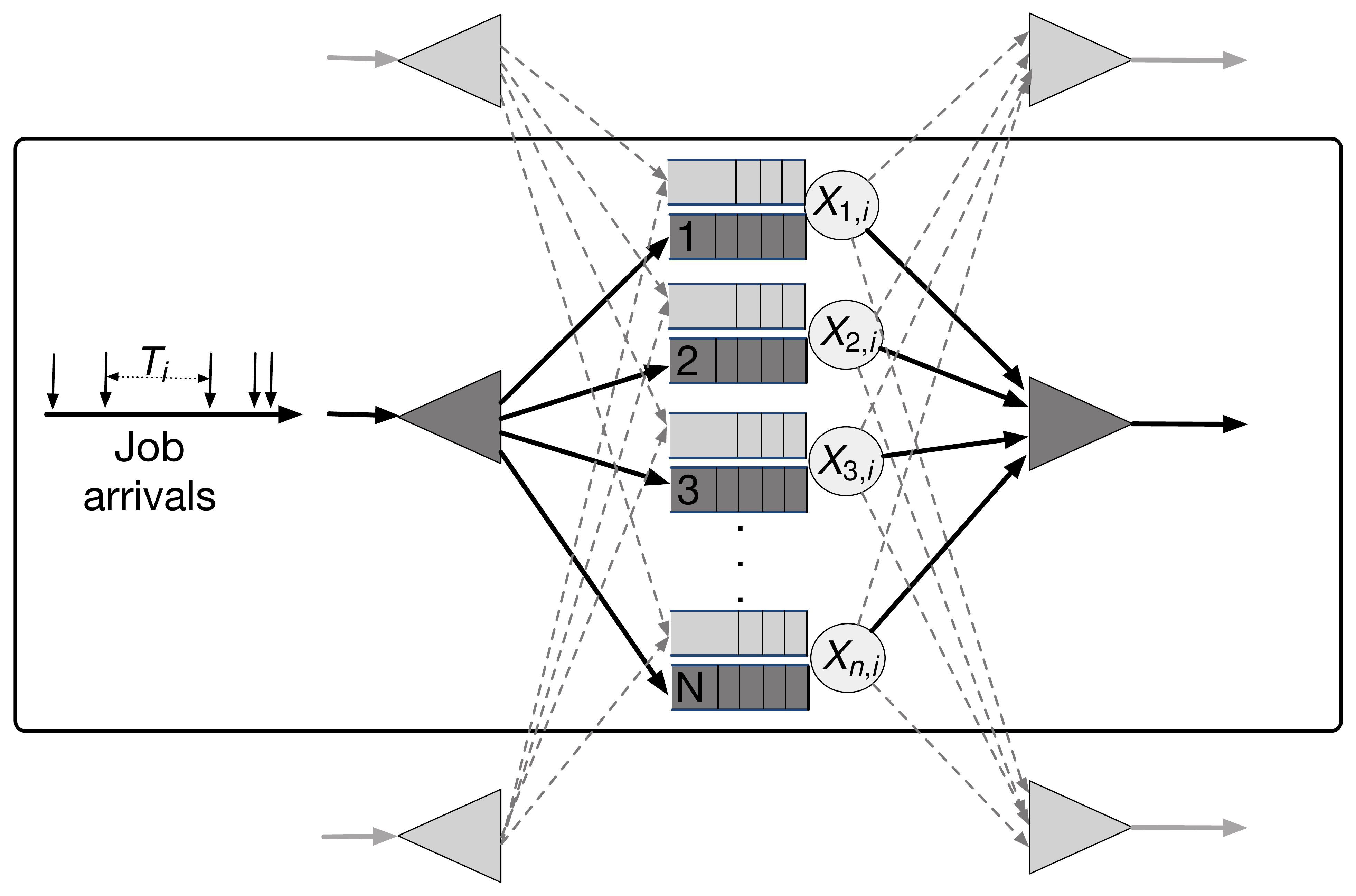}
	\caption{Single-node FJ system with shared resources.}
	\label{fig:shared_resources}
	\vspace{-3pt}
\end{figure}

\begin{figure*}[t!]
	\centering
	\begin{subfigure}[b]{0.32\textwidth}
		\centering
		\includegraphics[width=\textwidth]{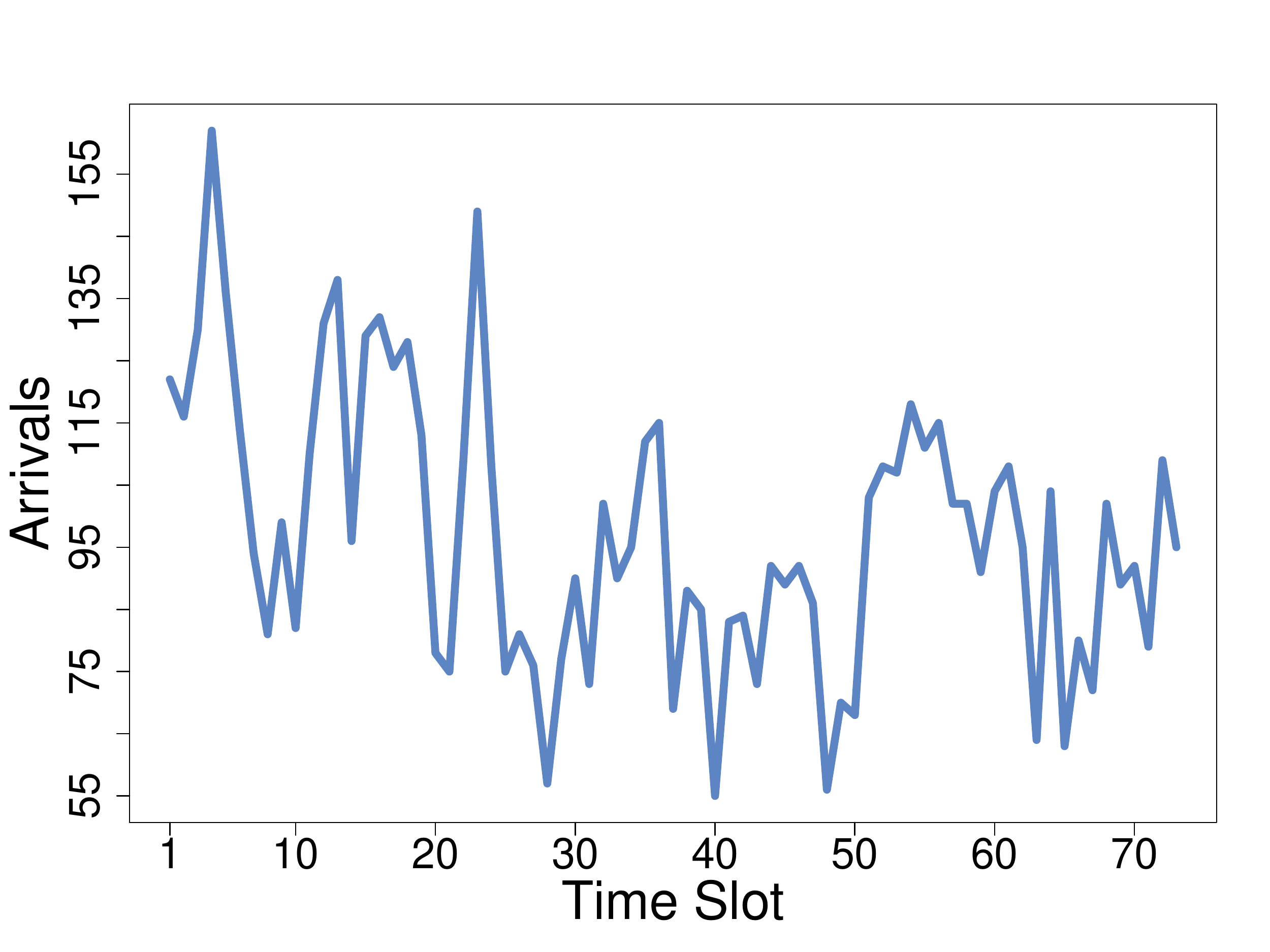}
	\end{subfigure}
	\begin{subfigure}[b]{0.32\textwidth}
		\centering
		\includegraphics[width=\textwidth]{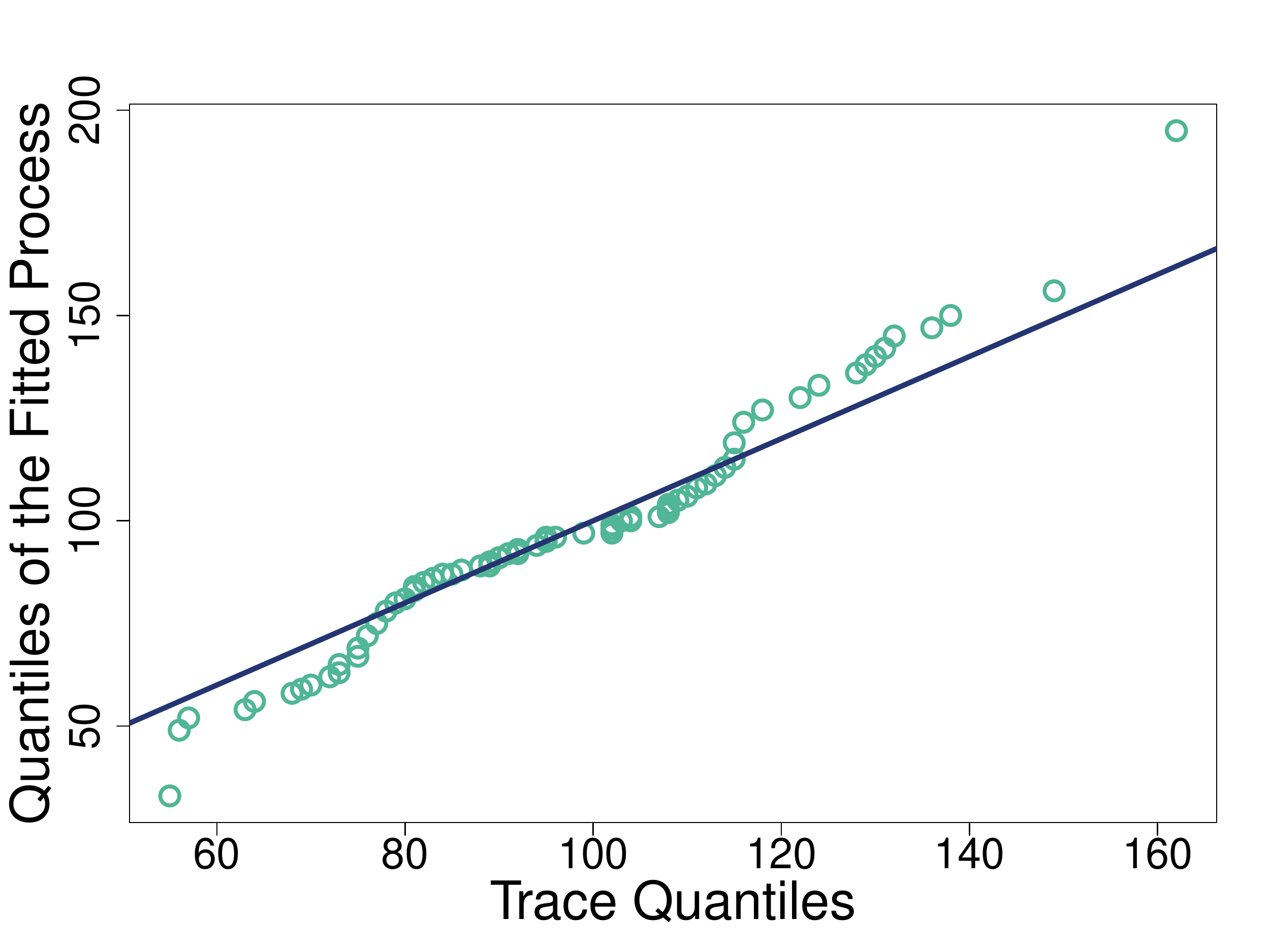}
	\end{subfigure}
	\begin{subfigure}[b]{0.32\textwidth}
		\centering
		\includegraphics[width=\textwidth]{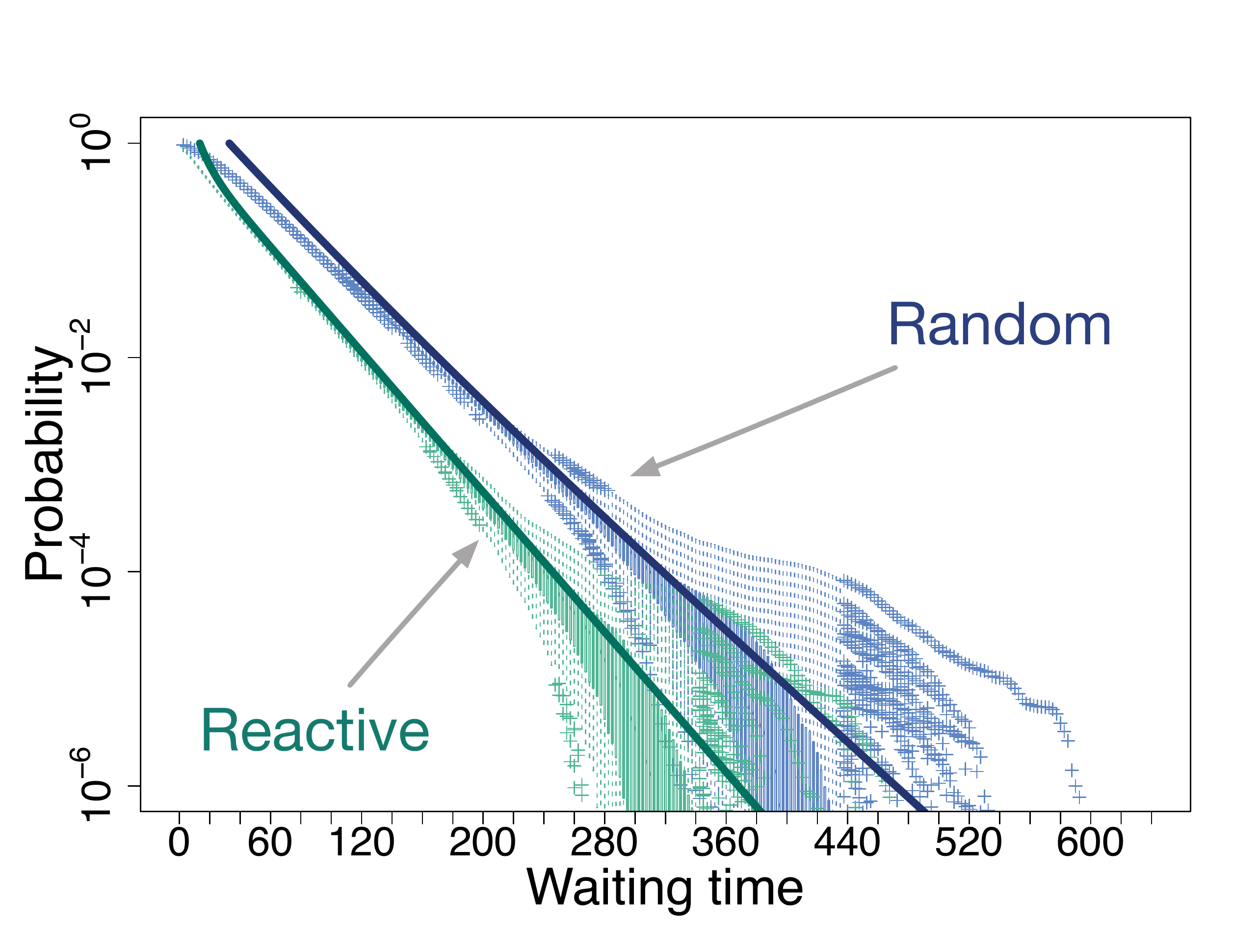}
	\end{subfigure}
	\caption{Numerical study based on real traces. \textbf{(Left)} Data trace plotted against observation time slots. Each time slot consists of $300$ seconds. 
		\textbf{(Middle)} Q-Q plot of data trace versus simulations of the fitted process. 
		\textbf{(Right)}  An FJ system with  the fitted arrival process   and with five servers with randomly assigned service rates. 
		A simple reactive provision is  designed and compared against the one assigning service rates randomly. The reactive provision rearranges the service rates of each of the servers 
		so as to have a high service rate when the arrival rate is high. }
	\label{fig:Google_Trace}
\end{figure*}

Here we characterize the arrival process from a real data trace and based on that,  construct a simple provision.




\subsection{Description of the dataset}
The data traces used in this discourse are available publicly  \cite{GoogleTrace}. It provides traces from a Borg cell (see \cite{Verma2015Borg}) recorded over a window of~$ 7$ hours. 
We assume a job arrives 
during the time slot where it first appears in the dataset. We  generate counts of arrivals in each time slot of $300$ seconds. To avoid adverse effects of outliers, first two time slots are discarded.

\subsection{Estimation Procedure}
Note that the data at our disposal do not contain inter-arrival times but rather counts of arrivals during time slots. 
Therefore, we formulate a Markov-modulated Poisson Process (MMPP) from which intensities of the inter-arrival times 
can be estimated. 
There are several estimation procedures 
for this purpose, including standard Maximum Likelihood Estimates (MLEs). The number of states being unknown a priori, one could also pose this as a (Bayesian) model selection problem. 
We employ the  ``LAMBDA'' algorithm proposed in \cite{heyman2003modeling}. 
The number of states  is estimated to be~$3$ and the per-second arrival intensities, $(\lambda_1, \lambda_2, \lambda_3)=(0.4616, 0.3180, 0.2011)$. The transition matrix is also estimated. 


\subsection{FJ System with a Reactive Provision}
We present an idealized situation to utilize the estimated parameters. Suppose we have five heterogeneous servers that are fed by Markov-modulated arrivals with the estimated transition probability matrix and inter-arrival intensities. Suppose each server can operate in three efficiency settings: high, medium, and low. In reality, the modulating chain is unobservable except perhaps for some special cases (\eg, distinguishable job types being represented by the chain). Therefore, to design a reactive provision, one needs to  estimate the hidden state from observable inter-arrival times. Machine learning techniques can be used to achieve this objective. But in light of \ref{itm:b1}, we do not attempt to do that here. For the sake of simple demonstration, we devise  a simple reactive provision assuming the  chain is visible. The provision simply assigns highest service rate when the arrival rate is highest, and assigns lowest, when the arrival rate is lowest. Mathematically, it just rearranges the service rates as described in Sec.~\ref{sec:everythingModulated} so that $\mu_{n,i} \geq \mu_{n,j}  $ whenever $i >j$ for all $n \in \setN{N}$, because the arrivals rates satisfy $\lambda_1>\lambda_2>\lambda_3$. This  provision is compared against a random assignment in Fig.~\ref{fig:Google_Trace}.


%% file: Sections/Discussion.tex
\section{Discussion and conclusions}
\label{sec:discussion}
In this paper, we have provided  computable upper bounds on  tail probabilities of  the steady-state waiting times for a general FJ system in   a Markov-additive process framework. We applied our results to three specific application areas and also presented an abstract conceptualization of     provisions. 
For the purpose of illustration, we also calibrated our model from real data traces.  

Our mathematical framework is a  general one. Most known cases can be derived by suitably choosing the kernels. For ease of understanding, we only provided simple examples involving the ``uncoupled'' 
case 
(see Fig.~\ref{fig:graphical_model}) in the preceding sections. In this closing section, we highlight the strength of our  model by mentioning two more ways in which it can be utilized.  

\paragraph*{Design of Proactive Provisions}
In Fig.~\ref{fig:graphical_model}, we mentioned that Markov-additive processes are  capable of modeling not only reactive 
but also proactive systems. As done in Sec.~\ref{sec:everythingModulated}, we model the changing environment by a Markov chain~$(C_k)_{k \in \setOfNonnegativeIntegers}$. However, in sharp contrast to reactive mechanisms in \ref{sec:everythingModulated}, we anticipate the immediate environmental changes and set service rates accordingly. Therefore, the distribution of the increments  $X_{n,k+1}^A$, for each $n \in \setN{N}$, and $ Y_{k+1}^A $  for the work-conserving  and blocking systems respectively, will also depend on $C_{k}$. Such provisions, we believe, will play a big role in the coming years of communication networking as it allow for preparedness and could potentially yield cost reduction. Owing to lack of space, we do not provide a detailed example here. 


\paragraph*{Renewal Processes}
We also point out that several previously known results on Fork-Join systems where a renewal arrival process was assumed (\eg, \cite{Rizk2015Sigmetrics,Rizk2016}) can also be retrieved by simply setting $\mathbb{E} = \{1\}$. In this case, following Algs.~\ref{alg:non_blocking} and \ref{alg:blocking}, the bounds turn out to be 
\begin{equation}
	\probOf{W \geq w } \leq \sum_{n \in \setN{N}} e^{- \theta_n w}  \eqkomma  \text{ and }	\probOf{W' \geq w } \leq e^{- \theta w} \eqkomma 
	\label{eq:renewal}
\end{equation}
where $ \theta_n = \sup \{ s>0 \mid  \Eof{ \myExp{ s S_{n,1}}  } \Eof{  \myExp{ -s A_1  }  } \leq 1 \} $ and $  \theta = \sup \{ s>0 \mid  \Eof{ \myExp{ s  \max_{n \in \setN{N}}   S_{n,1}}  } \Eof{  \myExp{ -s A_1  }  } \leq 1 \}   $. The bounds in~\eqref{eq:renewal} are generalizations of  \cite{Rizk2015Sigmetrics,Rizk2016} to heterogeneous servers. 

%


%% file: Sections/Appendix.tex
\appendices
\section{}
\label{sec:AppendixA}

\begin{myDefinition}
	\label{dfn:MA_process}  \textbf{(Markov-additive process)} The processes $\{(C_k, X_{n,k})  \}_{k \in \setOfNaturals}$, for each $n \in \setN{N}$, and $\{ (C_k,Y_k ) \}_{k \in \setOfNaturals}$  are Markov-additive (MA) processes on $(\mathbb{E} \times \setOfReals, \mathcal{E}\times \borel{\setOfReals}   )$ 
	if
	\begin{enumerate}
		\item The processes $\{ (C_k, X_{n,k}) \}_{k \in \setOfNaturals}$, for each $n \in \setN{N}$, and $\{ (C_k,Y_k ) \}_{k \in \setOfNaturals}$  are Markov processes on $(\mathbb{E} \times \setOfReals, \mathcal{E}\times \borel{\setOfReals}   )$.
		\item      The following holds for  $c \in \mathbb{E}, s \in \setOfReals, A \in \mathcal{E}, B \in \borel{\setOfReals}$, 
		\begin{align*}
		&	\probOf{ (C_{k+1}, X_{n,k+1}) \in A \times (B+s)  \mid (C_{1}, X_{n,1}) = (c,s)  } \\
		{}& =	\probOf{ (C_{k+1}, X_{n,k+1}) \in A \times B  \mid (C_{1}, X_{n,1}) = (c,0)  } 	\\
		{}& =	\probOf{ (C_{k+1}, X_{n,k+1}) \in A \times B  \mid C_{1}= c } 	 \eqkomma	\\
	\text{and } \;	&	\probOf{ (C_{k+1}, Y_{k+1}) \in A \times (B+s)  \mid (C_{1}, Y_{1}) = (c,s)  } \\
		{}& =	\probOf{ (C_{k+1}, Y_{k+1}) \in A \times B  \mid (C_{1}, Y_{1}) = (c,0)  } 	\\
		{}& =	\probOf{ (C_{k+1}, Y_{k+1}) \in A \times B  \mid C_{1} = c } 	 \eqpunkt 
		\end{align*}
		That is, we endow the Markov chain $\{C_k\}_{k \in \setOfNonnegativeIntegers}$  with additive components $\{ X_{n,k}\}_{n \in \setN{N}, k \in \setOfNonnegativeIntegers}$ and $\{Y_k\}_{k \in \setOfNonnegativeIntegers}$. Accordingly, define the transition kernels, for  $ n \in \setN{N}$,
		\begin{equation}
		\begin{aligned}
		K_n(c,   A \times B  ) \defeq {} & \probOf{ (C_{1}, X_{n,1}) \in A \times B  \mid C_{0} = c }   \eqkomma \\
		L (c,   A \times B  ) \defeq {} & \probOf{ (C_{1}, Y_{1}) \in A \times B  \mid C_{0} = c } \eqkomma
		\end{aligned} 
		\end{equation}
		where   $ c \in \mathbb{E}\, ,   A \in \mathcal{E}$ and $B \in \borel{\setOfReals}$.
	\end{enumerate}	
\end{myDefinition}

\paragraph*{Technical Assumptions}
We shall assume that conditional on $\{C_k= c\}$, the servers act independently. In addition to this, we make the following  assumptions
\begin{enumerate}[label=\color{tud0d}\textbf{B\arabic*}]
	\item \label{itm:as2}   \textbf{(Recurrence)} The process $\{C_k\}_{k \in \setOfNonnegativeIntegers}$  is  an aperiodic, irreducible Markov chain with respect to some maximal irreducibility measure and there exist probability measures $\iota_n$, for each $n \in \setN{N}$,  and $\nu$  on $(\mathbb{E} \times \setOfReals, \mathcal{E}\times \borel{\setOfReals}   )$, integers $m_{n,0}$, for each $n \in \setN{N}$, and $m_1$, and real numbers $0 < a_{n,0} \leq a_{n,1}  < \infty$, for each $n \in \setN{N}$, and $0 < b_0 \leq b_1 < \infty$ such that, for all $n \in \setN{N} $,
	\begin{align*}
	a_{n,0} \iota_n( A \times B ) \leq{}& K_n^{m_{n,0}} (x, A \times B)  \leq {} a_{n,1} \iota_n( A \times B )  \eqkomma \\
	b_{0} \nu( A \times B ) \leq{}& L^{m_{1}} (x, A \times B)  \leq{}  b_{1} \nu( A \times B )  \eqkomma 
	\end{align*}
	for each $ x \in \mathbb{E}\, ,   A \in \mathcal{E}$ and $B \in \borel{\setOfReals}$.
	\item \label{itm:as3} 	 \textbf{(Stability)} For stability of the system, we assume $\max_{n \in \setN{N}} \Eof{X_{n,1} } <0$ and $\Eof{Y_1 } <0 $. 
	\item \label{itm:as4} \textbf{(Finite cumulants)} Allowing possibly infinite values, define, for $s \in \setOfReals$, 	for each $n \in \setN{N}$, 
	\begin{align*}
	\lambda_k^{(n)} (s)  \defeq {} & k^{-1} \log \Eof{  \myExp{  s X_{n,k}  }  } \eqkomma \\
	\lambda^{(n)} (s)  \defeq {} &  \lim_{k \rightarrow \infty } k^{-1} \log \Eof{  \myExp{  s X_{n,k}  }  }  \eqkomma \\
\text{and } \;	\zeta_k(s)  \defeq {} & k^{-1} \log \Eof{  \myExp{  s Y_{k}  }  } \eqkomma \\
	\zeta (s)  \defeq {} &  \lim_{k \rightarrow \infty } k^{-1} \log \Eof{  \myExp{  s Y_{k}  }  }  \eqpunkt 
	\end{align*}
	To exclude pathological cases, we assume that the effective domains 
	of  $\lambda_k^{(n)}$ and $\lambda^{(n)}$,  and $	\zeta_k$ and $	\zeta$ include common open intervals containing~$0$.
\end{enumerate}


%% file: Sections/Appendix_B.tex
\section{}
\label{sec:AppendixB}
The proofs of Thms.~\ref{thm:nonblocking_general} and \ref{thm:blocking_general} follow \cite{Iscoe1985largeDev,Duffield1994}. However, for the sake of completeness we provide them here. The central idea is to construct suitable martingales for the additive components $X_{n,1}^A$ for $n \in \setN{N}$, and $Y_1^A$ by means of the large deviations properties and then use Doob's celebrated maximal inequality for (super)-martingales to get the bounds. 

\begin{proof}[Proof of Theorem~\ref{thm:nonblocking_general} ]
First define the cumulants
	\begin{align*}
\lambda_k^{(n)} (\theta)  \defeq {} & k^{-1} \log \Eof{  \myExp{  \theta X_{n,k}  }  } \eqkomma \\
\implies \lambda^{(n)} (\theta)  = {} &  \lim_{k \rightarrow \infty } \lambda_k^{(n)} (\theta)  \eqkomma
\end{align*}
for each $n \in \setN{N}$. In the light of \ref{itm:as2}, \ref{itm:as3}, and \ref{itm:as4}, the following statements are immediate
\begin{enumerate}[label=\color{tud0d}\textbf{C\arabic*}]
	\item \label{itm:c1} For all  $n \in \setN{N}$  and $\theta \in \effectiveDomain{\lambda^{(n)}}$,  $\myExp{ \lambda^{(n)}(\theta) }$ is the simple maximal eigenvalue of $\tilde{K}_n$.
	\item \label{itm:c2}  The corresponding right eigenfunction $\{ r_n(c, \theta); \,  c \in \mathbb{E}  \}$ satisfying 
	\begin{equation*}
	\myExp{ \lambda^{(n)}(\theta) } r_n(c, \theta)  = \int_{\setOfReals} \tilde{K}_n(c, d\tau ; \theta) r_n(\tau, \theta) \eqkomma
	\end{equation*}
	is positive and bounded above.
	\item \label{itm:c3} For all  $n \in \setN{N}$ , the functions $\lambda^{(n)} $ and $\lambda_k^{(n)}, \, k \in \setOfNaturals$ are both strictly convex and essentially smooth. 
	\item \label{itm:c4} Define the filtration
	\begin{align}
	\mathcal{F}_k \defeq \sigma( \{C_i\}_{i \in \setN{k}},  \{X_{n,i}  \}_{n \in \setN{N}, \, i \in \setN{k}}   ) \eqkomma
	\end{align}
	the $\sigma$-algebra generated by  the history of the process $\{ (C_k, X_{n,k}) \}_{k \in \setOfNaturals}$ till and including time point~$k$. Then, for each $n \in \setN{N}$, define 
	\begin{align}
	M_k^{(n)}(s) \defeq \myExp{ s X_{n,k} -k \lambda^{(n)} (s)  } r_n(C_k, s)  \eqpunkt
	\end{align}
	Then, $ 	M_k^{(n)}(s) $ is a martingale with respect to the filtration~ $	\mathcal{F}_k $.
\end{enumerate}

Please note that \ref{itm:c1} and \ref{itm:c2} are generalizations of the well known \emph{Perron-Frobenius} theorem for real matrices with positive entries. However, when the state space~$\mathbb{E}$ is not finite, one could still obtain similar results. The existence, and properties \ref{itm:c1} and  \ref{itm:c2} follow from \cite{Harris1963branching,Iscoe1985largeDev}.  The statements \ref{itm:c3} and \ref{itm:c4} are proved in \cite{Iscoe1985largeDev}. Also, see \cite{Duffield1994}. In the following, we would always normalize $r_n(., \theta)$ so that 
$\Eof{ r_n(C_0, \theta)   } =1$, for each $n \in \setN{N}$. 

Having constructed the martingales $ 	M_k^{(n)}(s) $, we can apply Doob's maximal inequality to obtain 
\begin{align}
\probOf{   \sup_{k \in \setOfNonnegativeIntegers} X_{n,k} \geq w } \leq  \phi_n( s ) \myExp{- s w} \eqkomma
\end{align}
for all $s \in \effectiveDomain{\lambda^{(n)}}$, following Theorem~$3$ of \cite{Duffield1994}. In particular, we get 
\begin{align}
\probOf{   \sup_{k \in \setOfNonnegativeIntegers} X_{n,k} \geq w } \leq  \phi_n( \theta_n ) \myExp{- \theta_n w} \eqkomma
\end{align} where $\theta_n \defeq \sup \{ s>0 \mid  \lambda^{(n)}(s) \leq 0 \}$ and $ \phi_n (s) \defeq \esssup \{ \indicator{ X_{n,1} >0   } / r_n( C_1, s)   \}$, after having normalized $r_n(., \theta)$ so that 
$\Eof{ r_n(C_0, \theta)   } =1$, for each $n \in \setN{N}$. The final bound is obtained as follows
\begin{align*}
	\probOf{W \geq w }  ={}& 	\probOf{  \sup_{k \in \setOfNonnegativeIntegers}   \sup_{n \in \setN{N}} X_{n,k}   \geq w }  \\
	={}& 	\probOf{  \sup_{n \in \setN{N}}  \sup_{k \in \setOfNonnegativeIntegers}  X_{n,k}   \geq w } \\
	\leq &    \sum_{ {n \in \setN{N}} }	\probOf{  \sup_{k \in \setOfNonnegativeIntegers}  X_{n,k}   \geq w } \\
	\leq &  \sum_{ {n \in \setN{N}} }  \phi_n( \theta_n ) \myExp{- \theta_n w} \eqpunkt 
\end{align*}
This completes the proof.
\end{proof}

\begin{proof}[Proof of Theorem~\ref{thm:blocking_general}]
First define the cumulants, 
		\begin{align*}
\zeta_k(s)  \defeq {} & k^{-1} \log \Eof{  \myExp{  s Y_{k}  }  } \eqkomma \\
\implies \zeta (s)  \defeq {} &  \lim_{k \rightarrow \infty } \zeta_k(s)  \eqpunkt 
\end{align*}
In the light of \ref{itm:as2}, \ref{itm:as3}, and \ref{itm:as4}, the following statements are immediate
\begin{enumerate}[label=\color{tud0d}\textbf{D\arabic*}]
	\item \label{itm:d1} For all $\theta \in \effectiveDomain{\zeta}$,  $\myExp{\zeta(\theta) }$ is the simple maximal eigenvalue of $\tilde{L}$. 
	\item \label{itm:d2}  The corresponding right eigenfunction $\{ r(c , \theta); \,  c \in \mathbb{E}  \}$ satisfying 
	\begin{equation*}
	\myExp{ \zeta(\theta) } r(c, \theta)  = \int_{\setOfReals} \tilde{L}(c, d\tau ; \theta) r(\tau, \theta) \eqkomma
	\end{equation*}
	is positive and bounded above.
	\item \label{itm:d3} The functions $\zeta $ and $\zeta_k, \, k \in \setOfNaturals$ are both strictly convex and essentially smooth. 
	\item \label{itm:d4} Define the filtration
	\begin{align}
	\mathcal{F}'_k \defeq \sigma( \{C_i\}_{i \in \setN{k}},  \{Y_{i}  \}_{ i \in \setN{k}}   ) \eqkomma
	\end{align}
	the $\sigma$-algebra generated by  the history of the process $\{ (C_k, Y_{k}) \}_{k \in \setOfNaturals}$ till and including time point~$k$. Then, for each $n \in \setN{N}$, define 
	\begin{align}
	M_k(s) \defeq \myExp{ s Y_{k} -k \zeta (s)  } r(C_k, s)  \eqpunkt
	\end{align}
	Then, $ 	M_k(s) $ is a martingale with respect to the filtration~ $	\mathcal{F}'_k $.
\end{enumerate}

Observe that, as before, \ref{itm:d1} and \ref{itm:d2} are generalizations of the well known \emph{Perron-Frobenius} theorem for real matrices with positive entries to uncountable state spaces~$\mathbb{E}$. 
The existence, and properties \ref{itm:d1} and  \ref{itm:d2} follow from \cite{Harris1963branching,Iscoe1985largeDev}.  The statements \ref{itm:d3} and \ref{itm:d4} are proved in \cite{Iscoe1985largeDev}. Also, see \cite{Duffield1994}. In the following, we would always normalize $r(., \theta)$ so that 
$\Eof{ r(C_0, \theta)   } =1$. 

After constructing the martingale $ 	M_k(s) $, we can apply Doob's maximal inequality to obtain 
\begin{align}
\probOf{   \sup_{k \in \setOfNonnegativeIntegers} Y_{k} \geq w } \leq  \phi( s ) \myExp{- s w} \eqkomma
\end{align}
for all $s \in \effectiveDomain{\zeta }$, following Theorem~$3$ of \cite{Duffield1994}. In particular, we get 
\begin{align}
\probOf{W \geq w }  ={} \probOf{   \sup_{k \in \setOfNonnegativeIntegers} Y_{k} \geq w } \leq  \phi( \theta ) \myExp{- \theta w} \eqkomma
\end{align} where $\theta \defeq \sup \{ s>0 \mid  \zeta(s) \leq 0 \}$ and $ \phi (s) \defeq \esssup \{ \indicator{ Y_{1} >0   } / r( C_1, s)   \}$ after having normalized $r(., \theta)$ so that 
$\Eof{ r(C_0, \theta)   } =1$.  This completes the proof.

\end{proof}

\begin{proof}[Proof of Remark~\ref{result:max-of-exponential}]
	The cumulative distribution function (CDF) of $Z$ is given by 
	\begin{eqnarray*}
		\probOf{V \leq z} &=& \prod_{i \in \setN{N}} (1- e^{-\mu_i z})  \eqkomma
	\end{eqnarray*}
	whence we derive the probability density function (pdf) of $Z$ as 
	\begin{eqnarray*}
		f_V(z) &=& \sum_{j \in \setN{N}} \mu_j e^{-\mu_j z} [ \prod_{i \in \setN{N} \setminus \{j\} } (1- e^{-\mu_i z})   ] \\
		&=& \sum_{j \in \setN{N}} \mu_j e^{-\mu_j z}  [   1+  \sum_{ \substack{ S \subset \setN{N} \setminus \{j \} \\ S \ne  \emptyset }} (-1)^{\cardinality{S}} \prod_{i \in S} e^{-\mu_i z}  ] \\
		&=& \sum_{j \in \setN{N}} \mu_j e^{-\mu_j z} [   1+  \sum_{ \substack{ S \subset \setN{N} \setminus \{j \} \\ S \ne  \emptyset } } (-1)^{\cardinality{S}}  e^{- z \sum_{i \in S} \mu_i }  ] \\
		&=& \sum_{j \in \setN{N}} \mu_j e^{-\mu_j z} [    \sum_{  S \subset \setN{N} \setminus \{j \} } (-1)^{\cardinality{S}}  e^{- z \sum_{i \in S} \mu_i }  ] \\
		&=&  \sum_{j \in \setN{N}} \mu_j  \sum_{ S \subset \setN{N} \setminus \{j \}  } (-1)^{\cardinality{S}}  e^{- z \sum_{i \in S \cup \{j\} } \mu_i } \\
		&=&\sum_{j \in \setN{N}} \mu_j  \sum_{ \substack{ S \subset \setN{N} \\ j \in S } } (-1)^{\cardinality{S}+1 }  e^{- z \sum_{i \in S } \mu_i }  \\
		&=&  \sum_{ \substack{ S \subset \setN{N}  \\ S \ne \emptyset } } (-1)^{\cardinality{S}+1 } (  \sum_{i \in S} \mu_i  ) e^{- z \sum_{i \in S } \mu_i }  \eqpunkt
	\end{eqnarray*}
	
	Therefore, the the moment generating function (MGF) of $Z$ is given by 
	\begin{align*}
		\Eof{e^{\theta V}} =& \int_{0}^{\infty} e^{\theta z} f_V(z)  \, dz \\ 
		=& \sum_{ \substack{ S \subset \setN{N}  \\ S \ne \emptyset } } (-1)^{\cardinality{S}+1 } \frac{  (  \sum_{i \in S} \mu_i  )    }{(  \sum_{i \in S} \mu_i  )  -\theta} \eqkomma
	\end{align*}
	and the mean of the distribution is given by 
	\begin{align*}
		\Eof{V} = &  \sum_{ \substack{ S \subset \setN{N}  \\ S \ne \emptyset } } (-1)^{\cardinality{S}+1 }  \frac{1  }{ (  \sum_{i \in S} \mu_i  ) } \eqpunkt 
	\end{align*}
	
%
	This completes the proof. 
	
\end{proof}